\documentclass[a4paper,12pt]{article}
\usepackage{jcappub}

\usepackage[greek,british]{babel}

\usepackage{latexsym,array,theorem,mathrsfs,subfigure,bm,float}
\usepackage{stackengine}
\usepackage{amsmath}
\usepackage{amsfonts}
\usepackage{amssymb}

\newcommand{\be}{\begin{eqnarray}}
\newcommand{\ee}{\end{eqnarray}}
\newcommand{\nn}{\nonumber}

\newcommand{\K}{{\rm K}}

\newcommand{\bl}{\textcolor{black}}

\newcommand{\bbl}{\textcolor{black}}

\newcommand{\obox}{\overset{(\Gamma)}{\Box}}
\newcommand{\onabla}{\overset{(\Gamma)}{\nabla}}

\newcommand{\oR}{\overset{(\Gamma)}{R}}

\newcommand{\gm}{\gamma_{\mbox{\tiny $X$}}}
\newcommand{\gmm}{\gamma_{\mbox{\tiny $XX$}}}

\newcommand{\PPsi}{{\psi}}

\newcommand{\gmp}{\gamma^\prime_{\mbox{\tiny $X$}}}

\newcommand{\km}{\kappa_{\mbox{\tiny $X$}}}
\newcommand{\kmm}{\kappa_{\mbox{\tiny $XX$}}}

\renewcommand{\theequation}{\arabic{section}.\arabic{equation}}

\begin{document}

\title{Varying the Horndeski Lagrangian  within the Palatini approach}


\author[a]{Thomas Helpin,}
\emailAdd{\tt thomas.helpin@lmpt.univ-tours.fr}

\author[b]{Mikhail~S.~Volkov}
\emailAdd{\tt michael.volkov@idpoisson.fr}

\affiliation[a,b]{
Institut Denis Poisson, UMR - CNRS 7013, \\ 
Universit\'{e} de Tours, Parc de Grandmont, 37200 Tours, France
}
\affiliation[b]{
Department of General Relativity and Gravitation, Institute of Physics,\\
Kazan Federal University, Kremlevskaya street 18, 420008 Kazan, Russia
}

\abstract{

\vspace{1 cm}
We analyse  what happens when the Horndeski Lagrangian is varied within the 
Palatini approach by considering the metric and connection as  independent variables. Assuming the connection to be 
torsionless,  there can be infinitely many metric-affine versions  $L_{\rm P}$  of the original Lagrangian which differ from 
each other by terms proportional to the non-metricity tensor. After integrating out the connection, 
each $L_{\rm P}$ defines a metric theory, which can either belong to the original Horndeski family,
or it can be of a more general DHOST type, or it shows the Ostrogradsky ghost. 
We analyse in detail the subclass of the  theory for which the equations are 
linear in the connection and find that its metric-affine version is ghost-free. \bbl{We present
a detailed classifications of homogeneous and isotropic cosmologies in these
theories.}
Taking into consideration other pieces of the Horndeski Lagrangian 
which are non-linear in the connection leads to more complex metric-affine theories which generically show the ghost.
 In some special cases the ghost can be removed by carefully adjusting the non-metricity contribution, but it is unclear 
if this is always possible. Therefore, the metric-affine generalisations of the Horndeski theory can be ghost-free, but not all of them are  ghost-free, 
neither are they  the only metric-affine theories for a gravity-coupled scalar field  which can be ghost-free.

}


\maketitle

\section{Introduction}
\setcounter{equation}{0}

The discovery of the cosmic acceleration  \cite{1538-3881-116-3-1009,0004-637X-517-2-565}
has invoked a large number of field-theory models of the Dark Energy.
 Most of them 
introduce a scalar field, as in the Brans-Dicke, quintessense, $k$-essence, etc. theories (see 
\cite{Copeland:2006wr,Joyce:2014kja} for reviews), while the others, 
as for example the $F(R)$ gravity \cite{Sotiriou:2008rp,DeFelice:2010aj}, although looking different, 
are equivalent to the theory with 
a scalar field.  Some of these models were actually introduced  long ago 
in the context of the inflation theory  \cite{Starobinsky:1980te}. 
In view of this interest towards theories with a gravitating scalar field one may ask:
what is the most general theory of this type described by second order equations of motion? 
The answer was obtained already in 1974 by 
Horndeski  \cite{Horndeski:1974wa} (and more recently rediscovered in \cite{Deffayet:2009wt,Deffayet:2011gz}): 
this theory 
is determined by the action 
\be                           \label{SH}
S_{\rm H}[g_{\mu\nu},\phi]=\int L_{\rm H}\, d^4x\,,
\ee
where, using the parameterization of   Ref.\cite{Kobayashi:2011nu},
\be                         \label{horn}
L_{\rm H}&=&({\cal L}_2+{\cal L}_3+{\cal L}_4+{\cal L}_5)\sqrt{-g},  \nonumber   \\            
{\cal L}_2&=&G_2(X,\phi)\,, \\
{\cal L}_3&=&G_3(X,\phi)\,[\hat{\Phi}]\,,\ \nonumber \\
{\cal L}_4&=&G_4({X},\phi)\,R-\partial_X G_4(X,\phi)\,
\left( [\hat{\Phi}]^2- [\hat{\Phi}^2]  \right) 
\,,  \nonumber \\
{\cal L}_5&=&G_5(X,\phi)\,[ \hat{G}\hat{\Phi}]
+\frac16\,\partial_X G_5(X,\phi)\,
\left(  [\hat{\Phi}]^3-3 [\hat{\Phi}] [\hat{\Phi}^2]+ 2[\hat{\Phi}^3]
\right).  \nn
\ee
Here $X=\frac12\,\partial_\mu\phi\partial^\mu\phi$ and 
$\hat{G}$, $\hat{\Phi}$ denote matrices  with components 
\be                        \label{phi00}
G^\mu_{~\nu}=R^\mu_{~\nu}-\frac12\, R\, \delta^\mu_{~\nu},~~~~~~~
\Phi^\alpha_{~\beta}=g^{\alpha\sigma}\nabla_\sigma\nabla_\beta\,\phi,
\ee
while  the 
brackets denote the trace, so that for example $[\hat{\Phi}]=\Box\phi$.
This theory incorporates  all previously studied models 
with a single gravity-coupled real scalar field.  
 The coefficient functions $G_k(X,\Phi)$ in \eqref{horn} can be arbitrary,  
 and  depending on their choice the properties
 of the theory can be different.

 There is a special subset of the theory, 
 sometimes call Kinetic Gravity Brading (KGB) theory \cite{Deffayet:2010qz}, \cite{Pujolas:2011he}, \cite{Easson:2011zy}
 defined by the following choice 
 of the coefficient functions: 
 \be                \label{KGB}
 G_4=G_4(\phi),~~~~G_5=0, 
 \ee
 while $G_2(\phi,X)$ and $G_3(\phi,X)$ can be arbitrary. The speciality of this choice  is that it defines 
 theories in which the 
 gravitational waves (GW)  propagate  with the speed of light 
 \bbl{around all backgrounds, as demonstrated in 
 \cite{Deffayet:2010qz}.}  If 
 the property \eqref{KGB} is not respected then the GW speed 
 is not constant and the corresponding theories are disfavoured  
 \cite{Creminelli:2017sry,Ezquiaga:2017ekz,Baker:2017hug,Sakstein:2017xjx}
 since the recent GW170817 event shows that the GW speed is equal to the speed of light with very high precision \cite{GW}. 
 
  The Horndeski theory can be generalised to the so-called DHOST models containing  
 higher order derivatives in the 
equations in such a way that the number of 
propagating degrees of freedom is still three \cite{Gleyzes:2014dya,Gleyzes:2014qga,Zumalacarregui:2013pma, 
 Langlois:2015cwa,
 Crisostomi:2016tcp
 }. 
 However, if one restricts only to theories with second order equations of motion, then the 
 Horndeski Lagrangian \eqref{horn}
 is the most general one to produce such theories
 within the {\it metric formulation}, that is assuming 
 the connection to be determined by the metric and the covariant derivative of the latter to vanish.

 In what follows we shall study theories obtained from the Horndeski 
 Lagrangian \eqref{horn} without imposing the metricity condition. 
   Specifically, we adopt 
  the Palatini approach  and vary the Lagrangian  independently with respect to 
 the metric $g_{\mu\nu}$, the scalar field $\phi$, and the connection  that  we assume to be {\it symmetric},  
 $\Gamma^\mu_{\alpha\beta}=\Gamma^\mu_{\beta\alpha}$.
 The equations for  
 $\Gamma^\mu_{\alpha\beta}$ are algebraic\footnote{\bbl{Unless $G_5(X,\phi)\neq 0$; see the remark after Eq.\eqref{KGB2}.}} 
 hence  the connection is non-dynamical, therefore  
 the number of propagating degrees of freedom is the same as in the original Horndeski theory, unless the ghost emerges.

 In fact,  we require  the connection to be symmetric just for simplicity, in order that the metric-affine theory 
 be maximally close to the original Horndeski theory. Relaxing  this condition would give  torsion-full generalisations of the theory, 
 and torsion should certainly be taken into account within the most-general metric-affine 
 setting \cite{Olmo:2013lta,Valdivia:2017sat, Bahamonde:2019shr}.  \bbl{At the same time,  the 
 vector part of the torsion 
 can  be gauged-away in projectively-invariant theories \cite{Afonso:2017bxr}, whereas requiring the theory to be 
 projectively invariant seems to be necessary for removing the ghost \cite{BeltranJimenez:2019acz}.
 In this work we shall mainly focus on the ${\cal L}_3$ part of the Horndeski theory, whose metric-affine version contains 
 the torsion in the vectorial form that can be gauged away. 
 As a result, choosing the connection to be torsionless from the very beginning does not actually  
 restrict the generality 
 of our analysis. This issue will be further discussed in \cite{PREP} (see also \cite{PREP1}). }

 The Ricci tensor in \eqref{horn}  is viewed as  function of 
 $\Gamma^\mu_{\alpha\beta}$, 
 \be             \label{Ricci}
R_{\mu\nu}\to \oR_{\mu\nu}\equiv \partial_\alpha \Gamma^\alpha_{\mu\nu}
-\partial_\nu \Gamma^\alpha_{\mu\alpha}
+\Gamma^\alpha_{\sigma\alpha}\Gamma^\sigma_{\mu\nu}
-\Gamma^\alpha_{\mu\sigma}\Gamma^\sigma_{\nu\alpha}\,,
\ee
the Ricci scalar and the Einstein tensor in \eqref{horn} are understood as 
\be                 \label{RG}
R\to \oR\equiv g^{\mu\nu}\oR_{\mu\nu},~~~~~G^\mu_{~\nu}\to  g^{\mu\sigma}\oR_{\sigma\nu}-\frac12 \oR\, \delta^\mu_{~\nu},
\ee
while the covariant derivatives should be computed with respect to   $\Gamma^\mu_{\alpha\beta}$, 
\be            \label{on}
\Phi^\alpha_{~\beta}\to g^{\alpha\sigma}\onabla_\sigma\onabla_\beta\,\phi. 
\ee
Making these replacements  in \eqref{horn},\eqref{phi00} gives us the metric-affine 
version of the original Horndeski  Lagrangian,
\be             \label{LP}
L_{\rm H}\to L_{\rm P},
\ee
and this defines the Palatini action 
\be                \label{SP}
S_{\rm P}[\Gamma^\sigma_{\alpha\beta},g_{\mu\nu},\phi]=\int L_{\rm P}\, d^4x\,.
\ee
Of course, this  reduces back to the original Horndeski action if the connection is set to be  Levi-Civita, 
\be    \label{back}
S_{\rm P}[\left\{^\alpha_{\mu\nu} \right\},g_{\mu\nu},\phi]=S_{\rm H}[g_{\mu\nu},\phi]. 
\ee
\bl{
Any additional matter
can be included into the theory by adding to the action the matter term 
$S_{\rm m}[g_{\mu\nu},{\cal X}]$, where ${\cal X}$ collectively denotes all matter fields. 
If the matter couples also to the connection and hence the matter term depends on 
$\Gamma^\sigma_{\alpha\beta}$ as well, this may generate a non-zero torsion. }

One should say that this metric-affine version of the original theory is not unique. 
 For example,  adopting instead of \eqref{on} the definition
 \be            \label{on1}
\Phi^\alpha_{~\beta}\to \onabla_\sigma( g^{\alpha\sigma}\onabla_\beta\,\phi)
\ee
 would give a different Lagrangian $\tilde{L}_{\rm P}$ and a different action $\tilde{S}_{\rm P}$
  that also reduce back to $L_{\rm H}$ and $S_{\rm H}$ 
 when   $\Gamma^\mu_{\alpha\beta}=\left\{^\alpha_{\mu\nu} \right\}$. However, 
varying $\tilde{S}_{\rm P}$ and ${S}_{\rm P}$  would  not give the same equations.  It is clear that the two definitions
\eqref{on} and \eqref{on1} differ from each other by the term containing 
the covariant derivative of the metric -- the non-metricity tensor 
\be
Q_{\alpha\mu\nu}\equiv \onabla_\alpha g_{\mu\nu}\,,
\ee
which does not vanish  in general. 
Using this tensor one can construct  generalisations of the 
 Lagrangian:
\be                         \label{LLLP}
L_{\rm P}\to \tilde{L}_{\rm P}=L_{\rm P}+\Delta L_{\rm P}\,,
\ee
where 
\be   \label{LLP}
\Delta L_{\rm P}=
(c_1\, Q^{\mu\alpha}_{~~\alpha}\nabla_\mu \phi
+c_2\,g_{\mu\nu} Q^{\mu\nu}_{~~\alpha}\nabla^\alpha \phi
+c_3\,Q^{\mu\nu}_{~~\alpha}\nabla_\mu\phi\nabla_\nu\phi \nabla^\alpha \phi+\ldots )\sqrt{-g}.
\ee
Here $c_1$, $c_2$, $c_3$ can depend on $X,\phi$ and the dots stand for all  possible terms 
containing higher powers of $Q^{\mu\nu}_{~~\alpha}$  and higher derivatives of $\phi$.  
As a result, there can be infinitely many different versions $\tilde{L}_{\rm P}$  of the Palatini Lagrangian. 
All of them coincide 
when the non-metricity vanishes, but otherwise they lead to different theories.  This ambiguity in defining  the 
theory may actually be important  for removing the ghost.
However, below we shall  be considering only the simplest version of the theory
for which it is sufficient to choose 
\be
\Delta L_{\rm P}=0. 
\ee
 More complex cases 
will be reported separately \cite{PREP} (see also \cite{PREP1}).

 Therefore, in what follows we shall vary  the Palatini action $S_{\rm P}[\Gamma^\sigma_{\alpha\beta},g_{\mu\nu},\phi]$ 
 defined by \eqref{SP}. 
 We do not expect to get the same equations as those obtained from the metric action 
 $S_{\rm H}[g_{\mu\nu},\phi]$, since already 
 for the $f(R)$ theory the metric formulation and Palatini formulation give different results \cite{Sotiriou:2008rp}. 
 The same is expected to happen also for the Horndeski theory.

 We find that the resulting theory obtained from $S_{\rm P}[\Gamma^\sigma_{\alpha\beta},g_{\mu\nu},\phi]$ 
 can show quite different properties, depending   on whether  the Lagrangian $L_{\rm P}$ 
 respects or not the KGB condition \eqref{KGB}. 
 If this condition is respected then 
 $Q_{\mu\nu\alpha}\neq 0$
 but    the non-metricity contributions  can be grouped into additional terms in the 
 effective energy-momentum tensor of the scalar field, and 
 the field equations can be represented in the form containing only 
 the ordinary metric covariant derivatives. 
 Remarkably, these equations turn out to be identically the same as those for a metric KGB theory corresponding 
 to a specific choice of the coefficients $G_2,G_3,G_4$ in the Lagrangian. Therefore, the Palatini approach 
 yields in this case a theory which is still in the same Horndeski  class.  
 
 \bl{
 This is a non-trivial result.  It can be traced to the fact  that 
 the connection in our theory is non-dynamical and can be integrated out, which 
 transforms the metric-affine theory to a {\it metric} theory containing  additional interactions 
  \cite{Delsate:2012ky,BeltranJimenez:2017doy}.  When the connection is integrated out, we obtain 
  a {metric} theory that is again of  Horndeski type but 
  with  additional  scalar-tensor interactions whose 
  structure, however,  may in general be not  the same as that in the Horndeski theory.  As a result, 
 this metric theory may  be outside the Hordeski family 
  and may contain ghost. It is therefore rather non-trivial that, 
  when the procedure is applied to the KGB theory, the ghost does not appear 
  and the additional interactions again have the KGB structure. One could say that the KGB family 
   is ``stable'' under the metric-affine treatment. 
  This,  together with the fact that the GW speed is constant,  confirms again the special status  of these  theories. 
  More general subsets of the Horndeski family are not stable in the same sense -- their metric-affine versions
  contain higher derivatives. }

 If the Lagrangian $L_{\rm P}$ does not respect the condition \eqref{KGB} then the equations contain 
  higher derivatives. In some cases they are not dangerous as 
 the theory turns out to be of the DHOST type, however  this is not always the case. For example, choosing 
 \be                \label{KGB1}
 G_4(X,\phi)\neq 0,~~~~G_2=G_3=G_5=0,
 \ee
 and $\Delta L_{\rm P}=0$ in \eqref{LLLP},  one finds that 
  the ghost is absent if $G_4=f(\phi)X$. 
\bbl{  In some other cases 
  it can be removed via adjusting  
 $\Delta L_{\rm P}\neq 0$ in \eqref{LLLP}  but it 
 remains unclear  if such a procedure always works. }
 
 The metric-affine versions of theories  with  
 \be                \label{KGB2}
 G_5(X,\phi)\neq 0
 \ee
 remain by far almost totally unexplored because \bbl{ the connection then becomes dynamical. 
 Specifically, the connection enters the action 
 algebraically, apart from the terms with $\oR_{\mu\nu}$.   Varying the latter with respect to the connection 
 produces result  of the form 
  $\Sigma^{\alpha\mu\nu}_\beta \onabla_\alpha\left( \delta  \Gamma^\beta_{\mu\nu}\right)$ 
  with $\Sigma^{\alpha\mu\nu}_\beta$  constructed 
   from $g_{\mu\nu}$  and derivatives of $\phi,X$. After integrating by parts, this gives rise to term
   $\onabla_\alpha\Sigma^{\alpha\mu\nu}_\beta $ in the equations. If $G_5=0$ then   
  $\Sigma^{\beta\mu\nu}_\alpha$ does not depend on the connection and 
   the resulting equations 
  are algebraic in $ \Gamma^\beta_{\mu\nu}$. 
 If $G_5\neq 0$ then  $\Sigma^{\beta\mu\nu}_\alpha$  contains 
 $\onabla_\mu\onabla_\nu\phi$ and therefore depends on  $ \Gamma^\beta_{\mu\nu}$ 
 hence  $\onabla_\alpha\Sigma^{\alpha\mu\nu}_\beta$ 
  contains derivatives $ \Gamma^\beta_{\mu\nu}$. 
   As a result, the equations for $ \Gamma^\beta_{\mu\nu}$  are algebraic if $G_5=0$ but they become differential if $G_5\neq 0$.  
   The connection starts propagating  
   in the latter case, which should considerably change the physical contents of the theory.  
    }

 The rest of this text is organised as follows. In Section \ref{SVar} we perform the Palatini variation of the piece of the 
 Lagrangian respecting the KGB condition \eqref{KGB}, and
 in Section \ref{SRel} we show that the resulting equations actually correspond to one of the metric Horndeski theories. 
 Therefore, varying the same KGB action in the metric approach and in the Palatini approach gives two different theories 
 from the same metric  KGB class. In Sections \ref{PM},\ref{CS} we study their solutions 
  to see how much these two theories differ from each other. 
   \bl{One should stress that, although both of these models are in the same KGB  class   and hence their tensor modes 
  propagate with the speed of light, the properties of the  scalar mode are not necessarily the same. 
  Cosmologies  in  these  two models are  not the same and their stability properties are different.} 
  \bl{In Section \ref{PM} we consider small perturbations of homogeneous and isotropic backgrounds, 
  which gives us conditions for the  absence 
 of ghosts and tachyons in the scalar sector.} In Section \ref{CS} we specify the subclass of theories  invariant 
 under shifts  $\phi\to\phi+\phi_0$ and describe  \bbl{all} homogeneous and isotropic
 cosmologies in these theories. \bbl{The  spectrum of these solutions is surprizingly  rich, and we make a comparison 
 with the solutions previously described  in the literature
 \cite{Deffayet:2010qz}, \cite{Pujolas:2011he}, \cite{Easson:2011zy}. }
  In Section \ref{SMore} we briefly describe what happens 
  if the condition \eqref{KGB} is not respected --  a more detailed analysis 
  will be reported separately \cite{PREP}. 
  We make some concluding remarks in Section \ref{SFin}. The Appendix contains the lengthy expression 
 for the connection arising in the $G_4(X,\phi)$-subset of the  theory. 

 The metric-affine formulation for the scalar-tensor theories was recently studied also  in \cite{Li:2012cc}, 
 \cite{Aoki:2018lwx}, \cite{Galtsov:2018xuc}, \cite{Shimada:2018lnm}, \cite{Aoki:2019rvi}. However, to the best of our knowledge, 
 the entire Horndeski family has not been systematically analysed  from this viewpoint.

\section{Varying the KGB part of the Palatini action \label{SVar}}
\setcounter{equation}{0}

Imposing  the KGB condition \eqref{KGB}, the action \eqref{SP} reduces to 
\be                  \label{action}
S_{\rm P}[ \Gamma^\alpha_{\mu\nu},g_{\mu\nu},\phi]=M_{\rm Pl}^2\int\left(G_4(\phi)\oR+K(\phi,X)+G_3(\phi,X) \obox\phi\right)\sqrt{-g}\, d^4 x\,, 
\ee
with the Ricci scalar  $\oR$ defined according to \eqref{Ricci},\eqref{RG}; \bbl{our signature is $(-,+,+,+)$}.
We assume the connection to be symmetric, 
$ \Gamma^\alpha_{\mu\nu}= \Gamma^\alpha_{\nu\mu}$, but the Ricci tensor $\oR_{\mu\nu}$ 
will not in general be symmetric, unless 
$ \Gamma^\alpha_{\mu\nu}$ is a Levi-Civita connection. 
The other quantities 
in the action are the squared gradient of the scalar field 
and covariant d'Alembertian, 
\be                \label{X}
X=\frac12\, g^{\mu\nu}\partial_\mu\phi\partial_\nu\phi\equiv g^{\mu\nu}X_{\mu\nu},~~~~
\obox\phi= g^{\mu\nu}\onabla_\mu\onabla_\nu\phi
=g^{\mu\nu}\left(\partial_{\mu\nu}\phi- \Gamma^\alpha_{\mu\nu}\partial_\alpha\phi\right).
\ee
Let us  vary the action \eqref{action} independently 
with respect to $ \Gamma^\alpha_{\mu\nu}$, $g_{\mu\nu}$, 
and $\phi$. 
To vary with respect to $\Gamma^\alpha_{\mu\nu}$,
 we notice that 
the only connection-dependent terms in the action 
are $\oR$ and $\obox\phi$. The variation 
 $ \delta \Gamma^\alpha_{\mu\nu}$ is a tensor that  induces 
 the  variations, 
 \be             \label{Ricci1}
\delta \oR_{\mu\nu}=\onabla_\alpha\left( \delta \Gamma^\alpha_{\mu\nu}\right)
-\onabla_\nu \left(\delta \Gamma^\alpha_{\mu\alpha}\right),~~~~~~~
\delta \obox\phi=-g^{\mu\nu}\partial_\alpha\phi \,\delta \Gamma^\alpha_{\mu\nu}\,.
\ee
Injecting this to \eqref{action}, integrating by parts and remembering that the metric 
is not necessarily covariantly constant with respect to $\onabla$, we obtain 
\be
\delta S_{\rm P}=\int \Delta^{\mu\nu}_{~~\alpha}\, \delta \Gamma^\alpha_{\mu\nu} \sqrt{-g}\, d^4x\,,
\ee
with 
\be
\Delta^{\mu\nu}_{~~\alpha}=\left.\left.\frac{1}{\sqrt{-g}}\, 
\onabla_\sigma \right(\sqrt{-g}\,G_4\, (\delta^{\mu}_\alpha\, g^{\nu \sigma } -\delta^\sigma_\alpha \,g^{\mu\nu})\right) 
 -G_3\, g^{\mu\nu}\partial_\alpha \phi\,.
\ee
The variation of the action will vanish if 
\be              \label{delta}
\Delta^{(\mu\nu)}_\alpha=0.
\ee
It follows that $\Delta^{(\mu\nu)}_\mu=0$, which yields 
\be              \label{del1}
\frac{1}{\sqrt{-g}}\, 
\onabla_\mu \left(\sqrt{-g}\, G_4\, g^{\mu\nu} \right)=\frac23\,G_3\, \partial^\nu\phi\,.
\ee
Taking this condition into account,  Eq.\eqref{delta}  reduces to 
\be              \label{del2}
\frac{1}{\sqrt{-g}}\, 
\onabla_\alpha \left(\sqrt{-g}\, G_4\, g^{\mu\nu} \right)=
G_3\left(
\frac23 \delta^{(\mu}_\alpha \partial^{\nu )}\phi - g^{\mu\nu}\partial_\alpha\phi
\right).
\ee
Since one has
\be             \label{gg}
\frac{1}{\sqrt{-g}} \onabla_\alpha \sqrt{-g}=-\frac12\, g_{\mu\nu} \onabla_\alpha g^{\mu\nu}\,,
\ee
one obtains after simple manipulations the following expression for the covariant derivative of the metric, 
\be
G_4\, \onabla_\alpha g^{\mu\nu}=  g^{\mu\nu} \partial_\alpha G_4+\frac23 G_3\left(
g^{\mu\nu}\partial_\alpha\phi+\delta^{(\mu}_\alpha \phi^{\nu)}
\right). 
\ee
This can be resolved to obtain the connection,
\be                   \label{Gamma1}
\Gamma^\alpha_{\mu\nu}=\left\{^\alpha_{\mu\nu}\right\}
+\frac12\left(\delta^\alpha_\mu \partial_\nu\omega + \delta^\alpha_\nu \partial_\mu\omega  
-g_{\mu\nu}\partial^\alpha\omega\right)+\frac13\,\gamma 
\left(\delta^\alpha_\mu \partial_\nu\phi + \delta^\alpha_\nu \partial_\mu\phi\right).
\ee
Here and in what follows we use the functions $\omega,\gamma,\kappa$ 
related to $G_4,G_3,K$ in the action via  
\be                       \label{def}
G_4=e^\omega,~~~~G_3=\gamma G_4, ~~~~~K=\kappa G_4. 
\ee
It is worth noting that the first and second terms on the right in \eqref{Gamma1} 
correspond to the Kristoffel symbols for the conformally related metric $\bar{g}_{\mu\nu}=e^\omega g_{\mu\nu}$. 
However, the last term  in \eqref{Gamma1} does not have the Levi-Civita structure. 

Injecting the expression for $\Gamma^\alpha_{\mu\nu}$ 
to \eqref{Ricci} gives the Ricci tensor,
\be                      \label{Ric}
\oR_{\mu\nu}&=&R_{\mu\nu}-\nabla_{\mu}\nabla_{\nu}\,\omega-\gamma\nabla_{\mu}\nabla_{\nu}\phi
-\frac12\, g_{\mu\nu}\left[
\Box\omega
+\partial_\sigma\omega\partial^\sigma\omega
+\gamma\,\partial_\sigma\omega\partial^\sigma\phi
\right]   \nonumber  \\
&&+\frac12\, \partial_\mu\omega \partial_\nu\omega+
\gamma \,\partial_{(\mu}\omega\partial_{\nu)}\phi+
\frac13\,\gamma^2\,\partial_\mu\phi\,\partial_\nu\phi -\partial_{(\mu}\gamma\partial_{\nu)}\phi   
 \nonumber  \\
&& +\frac53\, \partial_{[\mu}\gamma\partial_{\nu]}\phi\,,
\ee
where $R_{\mu\nu}$ and $\nabla_\mu$ are the standard 
Ricci tensor and covariant derivative constructed from $\left\{^\alpha_{\mu\nu}\right\}$ while $\Box=\nabla^\mu\nabla_\mu$. 
We note that the last term on the right in \eqref{Ric} is antisymmetric under $\mu\leftrightarrow\nu$.

Let us now vary the action with respect to $\phi$. One has 
\be
\delta X=\nabla^\mu\phi\nabla_\mu \delta\phi,~~~~~~
\delta\obox\phi=g^{\mu\nu}\onabla_\mu\onabla_\nu\delta\phi.
\ee
Injecting this to the action and integrating by parts yields 
\be
\delta S_{\rm P}=\int E_{\phi}\,\delta \phi\,\sqrt{-g}\,d^4x\,,
\ee
where 
\be                   \label{Ephi}
E_\phi&=&\partial_\phi G_4\,\oR+\partial_\phi K+\partial_\phi G_3\,\obox\phi \nn  \\
&&-\nabla_\mu\left(\partial_X K\,\nabla^\mu\phi \right)
-\nabla_\mu\left(\partial_X G_3\,\obox\phi\,\nabla^\mu\phi \right) \nn \\
&&+\frac{1}{\sqrt{-g}}\,\onabla_\mu\onabla_\nu \left(\sqrt{-g}\,G_3\, g^{\mu\nu} \right).
\ee
To compute the expression in the third line we 
set $G_3=\gamma G_4$, inject to \eqref{del1}, and  use \eqref{gg} to obtain 
\be              \label{del1a}
\onabla_\mu \left(\sqrt{-g}\, G_4\, g^{\mu\nu} \right)=
\sqrt{-g}\left(G_4\,\partial^\nu\gamma+\frac23 G_3\,\partial^\nu\phi\right).
\ee
Since for any vector $I^\mu$ one has 
\be                   \label{DIV}
\frac{1}{\sqrt{-g}}\onabla_\mu \left(\sqrt{-g}\,I^\mu \right)= \nabla_\mu I^\mu ,
\ee
it follows that 
\be
\frac{1}{\sqrt{-g}}\,\onabla_\mu\onabla_\nu \left(\sqrt{-g}\,G_3\, g^{\mu\nu} \right)=
\nabla_\mu \left(G_4\,\partial^\mu\gamma+\frac23 G_3\,\partial^\mu\phi\right). 
\ee
Collecting everything  together, the variation of the action with respect to the scalar field is 
\be
E_\phi\equiv -\nabla_\mu J^\mu +\Sigma\,,
\ee
with 
\be                 \label{JJ}
J^\mu&=&\left\{\partial_X K+B(2X\partial_X+1)G_3-\partial_\phi G_3 \right\}\partial^\mu\phi
+\partial_X G_3\left\{\Box\phi\, \partial^\mu\phi-\partial^\mu X\right\},  \nn \\
\Sigma&=&\partial_\phi K+\partial_\phi G_4 \oR+\partial_\phi G_3 \obox\phi\,.
\ee
Here and below the following two functions are used,
\be                   \label{B}
A=\omega^\prime\gamma+\frac32\,\omega^{\prime 2}-\frac13\,\gamma^2,~~~~~~~~~~
B=\omega^\prime -\frac23\gamma,
\ee
where the prime denotes differentiation with respect to $\phi$. 

Varying the action with respect to the metric is straightforward and yields 
\be
\delta S_{\rm P}=\int G_4 E_{\mu\nu}\,\delta g^{\mu\nu}\sqrt{-g}\, d^4 x\,,
\ee
where 
\be              \label{Einst}
E_{\mu\nu}&=&\oR_{\mu\nu}-\frac12 \oR g_{\mu\nu} +\gamma\onabla_\mu\onabla_\nu\phi \nn \\
&&+\left(\km+\gm\obox\phi\right) X_{\mu\nu}
-\frac12\left(\kappa+\gamma\obox\phi\right )g_{\mu\nu}\,,
\ee
with $\kappa$ defined in \eqref{def} and $\gm=\partial_X\gamma$, $\km=\partial_X\kappa$. 
The Ricci tensor $\oR_{\mu\nu}$   is given by  \eqref{Ric} and tracing it 
yields $\oR$. One has 
\be
\onabla_\mu\onabla_\nu\phi=\nabla_\mu\nabla_\nu\phi
-2\left(\omega^\prime+\frac23\gamma\right) X_{\mu\nu}+\omega^\prime X g_{\mu\nu}
\ee
with $X_{\mu\nu}$ defined in \eqref{X}, hence 
\be               \label{boxG}
\obox\phi=\Box\phi+2BX.
\ee
Summarizing the above discussion, the action will be stationary if $E_{(\mu\nu)}=0$
and $E_\phi=0$. This yields the field equations which can be rewritten solely in terms of the 
ordinary metric covariant derivatives. 
The $E_{(\mu\nu)}=0$ conditions reduce to 
\be                \label{eq1}
G_{\mu\nu}+T_{\mu\nu}=0\,,
\ee
where $G_{\mu\nu}$ is the  Einstein tensor for $g_{\mu\nu}$ while the effective 
energy-momentum tensor 
\be               \label{eq2}
T_{\mu\nu}&=&-\omega^\prime \partial_\mu\partial_\nu\phi-\gm\,
\partial_{(\mu}\phi\partial_{\nu)}X \\
&+&\left(\km+\gm \Box\phi 
-2\omega^{\prime\prime}-2\omega^{\prime 2}-2\gamma^\prime -2\gamma\omega^\prime
+2A+2BX\gm
\right)X_{\mu\nu}  \nn    \\
&+&\left(
\frac12\,\gm\,\partial_\sigma\phi\partial^\sigma X-\frac12\,\kappa
+\omega^\prime \Box\phi
+2\omega^{\prime\prime}+2\omega^{\prime 2}+\gamma^\prime+\gamma\omega^\prime-XA
\right)g_{\mu\nu}  \,,\nn
\ee
 with $A,B$ defined in \eqref{B}. 
The condition $E_\phi=0$ yields 
the equation for the scalar field,
\be               \label{eq3}
\nabla_\mu J^\mu=\Sigma\,,
\ee
 where $J^\mu,\Sigma$ are defined by \eqref{JJ} with $\obox\phi$ given by \eqref{boxG} and $\oR$
 obtained by tracing $\oR_{\mu\nu}$ in \eqref{Ric}. 
 A direct verification shows that  the differential consequence of 
 \eqref{eq1}, the covariant conservation condition, 
 \be
 \nabla^\mu T_{\mu\nu}=0,
 \ee
 indeed 
 follows from Eqs.\eqref{eq1}--\eqref{eq3}.

  \section{Relation to the metric version of the theory \label{SRel}}
 \setcounter{equation}{0}
 Let us now return to the action \eqref{action} and assume 
that $ \Gamma^\alpha_{\mu\nu}=\left\{^\alpha_{\mu\nu} \right\}$ is the 
Levi-Civita connection  determined by $g_{\mu\nu}$. The action then reduces to the Horndeski action
$S_{\rm H}[g_{\mu\nu},\phi]$. 
Varying  it with respect to $g_{\mu\nu}$ and $\phi$  yields the equations 
\be                \label{eq1a}
G_{\mu\nu}+T_{\mu\nu}=0\,,~~~~~~~~~~~\nabla_\mu J^\mu=\Sigma\,,
\ee
where 
\be               \label{eq2a}
T_{\mu\nu}&=&-\omega^\prime \partial_\mu\partial_\nu\phi-\gm\,
\partial_{(\mu}\phi\partial_{\nu)}X \\
&+&\left(\km+\gm \Box\phi 
-2\omega^{\prime\prime}-2\omega^{\prime 2}-2\gamma^\prime -2\gamma\omega^\prime
\right)X_{\mu\nu}  \nn    \\
&+&\left(
\frac12\,\gm\,\partial_\sigma\phi\partial^\sigma X-\frac12\,\kappa
+\omega^\prime \Box\phi
+2\omega^{\prime\prime}+2\omega^{\prime 2}+\gamma^\prime+\gamma\omega^\prime
\right)g_{\mu\nu}  \,,\nn    
\ee
and also 
\be               \label{JJa}
J^\mu&=&\left\{\partial_X K-\partial_\phi G_3 \right\}\partial^\mu\phi
+\partial_X G_3\left\{\Box\phi\, \partial^\mu\phi-\partial^\mu X\right\},  \nn \\
\Sigma&=&\partial_\phi K+\partial_\phi G_4 R+\partial_\phi G_3 \box\phi\,.
\ee
Surprisingly, a direct verification shows that  equations \eqref{eq1}, \eqref{eq2}, \eqref{eq3}, \eqref{JJ}  
of the metric-affine version  can be obtained 
from equations \eqref{eq1a}--\eqref{JJa}  of  the metric version by simply replacing  in the latter 
\be                   \label{kkappa}
\kappa\to \tilde{\kappa}=\kappa+2XA,
\ee
with $A$ given by \eqref{B}. Therefore, the Palatini theory derived from the action \eqref{action}
is actually equivalent to the metric theory derived from the action 
\be                  \label{act1}
\tilde{S}_{\rm H}[g_{\mu\nu},\phi]=M_{\rm Pl}^2\int\left\{G_4(\phi) R+\tilde{K}(\phi,X)+G_3(\phi,X) \Box\phi\right\}\sqrt{-g}\, d^4 x\,,
\ee
with the new k-essence term 
\be                                   \label{KKK}
\tilde{K}=\tilde{\kappa} G_4=K+2XG_4 A=K+\left( 2G_3\partial_\phi G_4 +3 (\partial_\phi G_4)^2-\frac23 G_3^2 \right)\frac{X}{G_4}\,.
\ee
The explanation of this  is as follows. Let us return to the Palatini 
action \eqref{action} and inject into it the on-shell value of the connection, 
\be
\Gamma^\sigma_{\rho\gamma}=\Gamma^\sigma_{\rho\gamma}\left(g_{\alpha\beta},\phi\right), 
\ee
given by \eqref{Gamma1}. Using $\oR_{\mu\nu}$ and $\obox\phi$ expressed by \eqref{Ric} and 
\eqref{boxG} then yields 
\be
S_{\rm P}[\Gamma^\sigma_{\rho\gamma}\left(g_{\alpha\beta},\phi\right),g_{\mu\nu},\phi]=
\tilde{S}_{\rm H}[g_{\mu\nu},\phi],
\ee
so that the metric action \eqref{act1} is indeed recovered. \bl{This implies that 
the equations derived from both actions should coincide.}
Indeed, 
let us vary the scalar field, $\phi\to \phi+\delta \phi$. This induces the 
variations 
\be
\delta S_{\rm P}=\frac{\delta S_{\rm P}}{\delta \Gamma^\sigma_{\rho\gamma} }\frac{\partial \Gamma^\sigma_{\rho\gamma}\left(g_{\alpha\beta},\phi\right) }{\partial \phi}\,\delta \phi
+\frac{\delta S_{\rm P}}{\delta \phi}\,\delta \phi=\delta\tilde{S}_{\rm H}=\frac{\delta \tilde{S}_{\rm H}}{\delta \phi}\,\delta \phi\,. 
\ee
Since the connection is assumed to have the on-shell value, one has 
\be
\frac{\delta S_{\rm P}}{\delta \Gamma^\sigma_{\rho\gamma} }=0,
\ee
therefore 
\be
\frac{\delta S_{\rm P}}{\delta \phi}=\frac{\delta \tilde{S}_{\rm H}}{\delta \phi}\,,
\ee
hence the scalar field equation derived from the Palatini action $S_{\rm P}$ coincides with the one obtained from the metric action 
$\tilde{S}_{\rm H}$. The same applies for equations obtained by varying the metric, hence 
 theories derived from the Palatini action \eqref{action} and from the metric action 
\eqref{act1} are equivalent.  A similar equivalence holds for all other Horndeski models with $G_5=0$ as well,  because 
a non-dynamical connection 
 can always be integrated out and  the metric-affine theory reduces to a metric theory.

Summarizing, varying the same action \eqref{action} within the metric approach and within the Palatini approach 
yields two different theories from the same metric KGB class. In the former case one obtains the theory 
with coefficient functions $G_3$, $G_4$, $K$ while in the latter case one obtains theory with coefficients 
$G_3,G_4,\tilde{K}$, with $\tilde{K}$ defined by \eqref{KK}. Both theories are ghost-free 
and the GW speed is equal to one. 
Below we shall study solutions of these 
two theories  to see how much they differ from each other.

\bbl{
The change $K\to\tilde{K}$ expressed by \eqref{KKK}  can be viewed as a ``duality relation'' between different theories, 
and one can look for its interpretation, for example within the effective hydrodynamical description developed in  Ref.\cite{Pujolas:2011he}. 
If the theory is invariant under shifts $\phi\to \phi+\phi_0$ then the current 
in \eqref{JJa}  is conserved, which can be expressed as 
\be                       \label{diffusion}
\nabla_\mu (n\, u^\mu)=\nabla_\mu({\bm\kappa} a^\mu), 
\ee
 with the ``fluid 4-velocity" $u_\mu= \partial_\mu\phi/m$,
``acceleration" $a_\mu=u^\alpha\nabla_\alpha u_\mu$,  ``chemical potential" $m=\sqrt{2|X|}$,  ``density" 
$n=m\partial_X K+{\bm \kappa} \theta$ where the ``expansion" $\theta=\nabla_\alpha u^\alpha$, and with the 
``diffusivity" ${\bm \kappa}=2X\partial_X G_3$ (not to be confused with our $\kappa=K/G_4$).   
The  right-hand-side in \eqref{diffusion} can be interpreted as the ``diffusion term'' \cite{Pujolas:2011he}. 
}

\bbl{
Now, \eqref{KKK}  does not change $G_3$ and  the diffusivity ${\bm \kappa}$, but it changes $K$ (equation of state) 
and the density $n$. If repeated many times, it changes $K$ more and more, but 
it becomes identity if the theory is 
chosen such that $A=0$ and hence $\tilde{K}=K$. Unfortunately, this  condition does not have non-trivial solutions
in the shift-invariant theory\footnote{\bbl{The relation $\tilde{K}=K$ can be realized in a non-trivial 
way if the theory includes  non-metricity terms ${\Delta L}_{\rm P}$ of the type \eqref{LLP} \cite{PREP1}. }}. The duality  does not change
the right-hand-side of \eqref{diffusion}, but  since $n$ changes, the whole diffusion is affected. 
Eq.\eqref{diffusion} assumes the standard diffusion form in the limit of small $n$ and $m$, with the diffusion 
coefficient ${\cal D}=-{\bm\kappa}/(m\partial_m n)$  \cite{Pujolas:2011he}, but since the duality only affects  higher powers of $m$, 
${\cal D}$ remains invariant in this limit. 
The impact of the duality on the diffusion away from the weak field limit should probably be analyzed  
separately. 
 }

 \section{Stability of the scalar sector \label{PM}}
 \setcounter{equation}{0}
 
 \bl{
 In Section \ref{CS} we shall study cosmologies in  the Palatini-derived KGB  theory  and 
compare them with the those in the metric-derived theory.  Even though these two theories belong to the same metric KGB class,
hence they are free from  Ostrogradsky ghost and contain three propagating degrees of freedom, their scalar sector
can  be unstable. 
  We therefore study in this Section
small perturbations of cosmological backgrounds. 
Decomposing the perturbations  into tensor and scalar parts, one finds, as expected, 
that tensor modes always propagate with the speed of light. However, depending on the background, the scalar sector may contain 
ghost and tachyon instabilities. 
Their  absence  requires positivity of the kinetic coefficient $\K$ and 
sound speed squared $c_s^2$ derived in Eqs.\eqref{K},\eqref{XXX}. These conditions will be analysed in Section VI.}

 
 Let us assume the spacetime metric to be homogeneous and isotropic,
 \be                      \label{hom1}
 ds^2=-dt^2 + a(t)^2 \left( dx^2+dy^2+dz^2\right),
 \ee 
 while the scalar field to depend only on time,
 \be                      \label{hom2}
 \phi=\phi(t),~~~~~\PPsi\equiv \dot{\phi}.
 \ee
 The Einstein equations \eqref{eq1} of the Palatini version of the theory reduce to 
 \be             \label{eqs}
 3H^2=&-&\frac12\,\kappa+\frac32 \,(\gm\,\PPsi^2-2\omega^\prime)\,\PPsi \, H \nn \\
 &+&\frac12\left(\gamma^\prime+\frac13\gamma^2-\frac32\omega^{\prime 2}-\km \right) 
 \PPsi^2
 +\frac16\,(3\omega^\prime -2\gamma)\gm\,\PPsi^4  \nn  \,, \\
 2\dot{H}=&&\left(\frac12\gm\, \PPsi^2-\omega^\prime \right)\dot{\PPsi}   \nn \\
 &-&\left(\omega^{\prime\prime}+\frac14\omega^{\prime 2}+\frac12\gamma^\prime+\frac16\gamma^2 \right)\PPsi^2
 -3H^2
 -2\omega^\prime H\PPsi-\frac12\kappa\,,
 \ee
 whose consequence is the scalar field equation \eqref{eq3}. 
 Here  $H=\dot{a}/a$ and the prime denotes  differentiation with respect to $\phi$. 
  
 Suppose one finds a solution of Eqs.\eqref{eqs} (examples will be given below)
 describing a homogeneous and isotropic background \eqref{hom1},\eqref{hom2}. 
 Consider small perturbations of this background, 
 \be                     \label{pert0}
 g_{\mu\nu}\to g_{\mu\nu}+\delta g_{\mu\nu},~~~~~~~\phi\to \phi+\delta\phi.
 \ee
 In the linear approximation, the perturbations fulfill the equations obtained by perturbing the 
 background equations, 
 \be           \label{pert}
 \delta E_{(\mu\nu)}=0,~~~~\delta E_\phi=0. 
 \ee
The metric perturbations can be decomposed into the 
 scalar, vector, and tensor parts via 
 \be      \label{h}
 \delta g_{00}&=&- {\rm S}_3,   \nn \\
 \delta g_{0i}&=&a \left( \partial_i {\rm S}_4+{\rm W}_i  \right),   \nonumber \\
 \delta g_{ik}&=&a^2\left(  
 {\rm S}_1\,\delta_{ik}+\partial^2_{ik}{\rm S}_2+\partial_i {\rm V}_k+\partial_k {\rm V}_i+{\rm D} _{ik}
 \right),
 \ee
 where 
 \be
\sum_k \partial_k {\rm V}_k=\sum_k \partial_k {\rm W}_k=0,~~~~~
\sum_k \partial_k {\rm D}_{ki}=0,~~~~\sum_k {\rm D}_{kk}=0.
 \ee
 The spatial dependence is given by the plane waves where  the wave vector 
 can be oriented along the z-axis, so that the scalar modes are 
 \be
 {\rm S}_1=S_1(t) e^{ipz},~
{\rm S}_2=S_2(t) e^{ipz},~
{\rm S}_3=S_3(t) e^{ipz},~
{\rm S}_4=S_4(t) e^{ipz},~
\delta\phi=f(t) e^{ipz},~
 \ee
 the vector amplitudes are chosen as 
 \be
 {\rm V}_k=\,[V_{1}(t),V_{2}(t),0]\,e^{ipz},~~~~~~~
 {\rm W}_k=[W_{1}(t),W_{2}(t),0]\,e^{ipz},
 \ee
 while for the tensor modes the only non-trivial components of ${\rm D}_{ik}$ are 
 \be
 {\rm D}_{11}=-{\rm D}_{22}=D_{1}(t)\,e^{ipz},~~~~~
 {\rm D}_{12}={\rm D}_{21}=D_{2}(t)\,e^{ipz}. 
 \ee
Inserting everything into the perturbation equations \eqref{pert}  
 splits them  into three independent groups for the scalar, vector, and tensor modes. These equations 
determine the effective action which also spits into  three independent terms, 
 \be                  \label{act0}
 I\equiv I_{\rm T}+I_{\rm V}+I_{\rm S}=\frac{M_{\rm Pl}^2}{2} \int  \left(\delta E_{\mu\nu}\,\bar{\delta g}^{\mu\nu}
 +\delta E_\phi \,\bar{\delta\phi} \right) \, a^3\, d^4x , 
 \ee
 where the bar denotes complex conjugation. 
 One obtains in the tensor sector 
 \be                 \label{IT}
 I_{\rm T}=\frac{M_{\rm Pl}^2}{2}\, \int \K \left(  
 \dot{D}_{1}^2+\dot{D}_{2}^2- c_s^2\,\frac{p^2}{a^2}  (D_{1}^2+D_{2}^2)
   \right)\, a^3\,d^4x,
 \ee
 where the kinetic coefficient 
 $
 \K=G_4=e^\omega
 $
 (not to be confused with $K$) 
is always positive while the sound speed $c_s=1$. Therefore,
the gravity waves propagate with the speed of light as expected. 

The analysis in the vector sector shows that the vector modes have no kinetic term
and $I_{\rm V}=0$,  hence vector modes do not propagate.  

The analysis in the scalar sector is more involved but facilitated by the fact that one can impose the gauge where 
$\delta\phi=0$ (unless  $\PPsi=0$).  The equations then imply that the scalar amplitudes 
$S_2$, $S_3$, $S_4$ can be expressed in terms of $S_1$ and the effective action reduces to 
\be                 \label{ITs}
 I_{\rm S}=\frac{M_{\rm Pl}^2}{2}\, \int {\K}\left(  
\dot{S}_{1}^2-c_s^2\,\frac{p^2}{a^2} \, S_{1}^2
   \right)\, a^3\,d^4x\,,
 \ee
 with 
 \be                    \label{K}
 \K=\frac{G_4\PPsi^2}{6W^2}\,\Delta_1\,,~~~~~~c_s^2=\frac{\Delta_2}{\Delta_1}\,,
 \ee
 where 
 \be               \label{XXX}
 \Delta_1&=&(17\gm^2-12\omega^\prime \gmm+8\gamma\gmm)\PPsi^4-36H\gmm\PPsi^3 \nn   \\
 &&+(12\omega^\prime\gm-40\gamma\gm-12\gmp+12\kmm)\PPsi^2 \nn \\ 
 &&+72H\gm\PPsi  +8\gamma^2-12\km+24\gamma^\prime \,, \nn \\
\Delta_2&=&-3\gm^2\PPsi^4+(12\omega^\prime\gm-8\gamma\gm+12\gmp)\PPsi^2\nn \\
 &&+48H\gm\PPsi+(24\gm-12\gmm\PPsi^2)\dot{\PPsi}+8\gamma^2-12\km+24\gamma^\prime\,,  \nn \\
 W&=&4H+2\omega^\prime \PPsi-\gm \PPsi^3.
 \ee
 Both the kinetic term $\K$ and  sound speed squared $c_s^2$ should be positive for the system to be stable. 
Summarizing, the theory shows two propagating modes in the tensor sector and one scalar mode. 
The tensor modes  propagate with the speed of light, as expected, while  properties of the scalar mode 
depend on the background. The above formulas apply for the Palatini-derived theory described by 
\eqref{eq1}, \eqref{eq2}, \eqref{eq3}, \eqref{JJ}.  
The corresponding formulas in the metric theory described by 
\eqref{eq1a}--\eqref{JJa} are obtained by making in \eqref{eqs}, \eqref{XXX} the 
inverse to \eqref{kkappa} replacement : $\kappa\to \kappa-2XA$. 

\section{Cosmologies  \label{CS}}
\setcounter{equation}{0}


In order to study concrete  solutions, 
we must specify  the 
functions $G_4(\phi)$, $G_3(\phi,X)$, $K(\phi,X)$. 
We assume them 
to be independent of $\phi$, 
\be
G_4=const.,~~~~~G_3=G_3(X),~~~~~~K=K(X),
\ee
in which case the  theory is invariant under shifts 
\be
\phi\to\phi+\phi_0.
\ee
As the simplest option, we assume $G_3$ and $K$ to be  linear in $X$, hence 
\be              \label{AM}
G_4=1,~~~~G_3=\gamma=\alpha X,~~~~~K=\kappa=\beta X-2\Lambda,
\ee
where $\alpha,\beta,\Lambda$ are constant parameters, so that 
\be
\gm=\alpha,~~~~~\km=\beta. 
\ee
Eqs.\eqref{eq1}   then become 
\be      \label{T1}
G_{\mu\nu}+T_{\mu\nu}&=&0, ~
\ee
with the energy-momentum tensor 
\be                           \label{Ten}
T_{\mu\nu}=&-&\alpha \,\partial_{(\mu}\phi\partial_{\nu)}X+(\beta+\alpha\Box\phi-2\alpha^2 X^2)X_{\mu\nu}  \nn \\
&+&\left(\Lambda-\frac12\beta X+\frac12\,\alpha\, \partial_\sigma\phi \partial^\sigma X+\frac13\,\alpha^2 X^3 \right)g_{\mu\nu}\,,
\ee
while the scalar field equation \eqref{eq3} becomes total derivative, 
\be          \label{J1}
\nabla_\sigma J^\sigma&=&0,
\ee
with the current 
\be                \label{cur} 
J^\mu=(\beta-2\alpha^2 X^2)\partial^\mu\phi+\alpha\, (\Box\phi\partial^\mu\phi-\partial^\mu X). 
\ee
The equations of the corresponding metric version of the theory are obtained by simply omitting 
in \eqref{Ten},\eqref{cur} the terms proportional to $\alpha^2$.

\bl{
We shall now study  homogeneous and isotropic cosmologies of this shift-invariant theory, 
first within its Palatini version and then within its metric version. 
Even though these two  theories  belong to the same KGB class, their solutions are different
since the equations are not the same  and the stability conditions 
derived in Section~\ref{PM}
are also not the same, since they are background-dependent.}

\bbl{
It turns out that  the problem reduces to a single algebraic equation \eqref{e1}  which 
determines algebraic curves  whose position and critical points determine
the  properties of the cosmologies. One finds in this way many  different solutions: 
self-accelerating cosmologies, recollapsing cosmologies, and bounces. 
We give  below their  detailed  classification, 
analyse their stability conditions, identify stable branches, 
and make comparison  with the previously known results 
\cite{Deffayet:2010qz}, \cite{Easson:2011zy}}. 

\subsection{Master equations}

Assuming  the  homogeneous and isotropic ansatz \eqref{hom1},\eqref{hom2} for the fields, 
the Einstein equations \eqref{T1} reduce to 
\be      \label{Fr}
3H^2&=&\frac32\,\alpha\PPsi^3 H-\frac{1}{4}\,\beta\,\PPsi^2+\frac{5}{24}\alpha^2 \PPsi^6+\Lambda,  \\
2\dot{H}+3H^2&=&\frac12\,\alpha\,\PPsi^2\dot{\PPsi}+\frac14\,\beta\,\PPsi^2-\frac{1}{24}\,\alpha^2\, \PPsi^6+\Lambda\,,
\label{FF}
\ee
with $\PPsi=\dot{\phi}$. 
These equations can also be obtained by injecting \eqref{AM} to \eqref{eqs}. 
The only non-trivial component of the scalar current \eqref{cur}   is 
\be                      \label{Jzero}
J^0=\left(\beta-3\alpha H \PPsi-\frac12\,\alpha^2 \PPsi^4  \right)\PPsi\,,
\ee
and the scalar field equation \eqref{J1} reads 
\be
\frac{d}{dt}\left( a^3 J^0\right)=0, 
\ee
which implies that 
\be            \label{JJJ}
J^0=\frac{C}{a^3}\,,
\ee
where $C$ is the integration constant -- the scalar charge. 

The simplest solution of these equations is $C=\PPsi=0$ and $3H^2=\Lambda$. 
This solution is stable, although the general stability 
analysis carried out above does not apply in this particular case since the gauge $\delta\phi=0$ 
cannot be imposed if $\PPsi=0$. One should repeat the analysis keeping $\delta\phi\neq 0$
and then one finds in the scalar sector  $\K=c_s^2=1$. 

For solutions with $\PPsi\neq 0$ one can use the general formulas   \eqref{K} for the 
kinetic term and the sound speed, which  now reduce to 
\be             \label{KK}
\K=\frac{\PPsi^2(13\alpha^2\PPsi^4+24\alpha H\PPsi-4\beta)}{2(\alpha\PPsi^3-4H)^2},~~~~~~
c_s^2=\frac{\alpha^2\PPsi^4+16\alpha H\PPsi+8\alpha\dot{\PPsi}-4\beta}{13\alpha^2\PPsi^4+24\alpha H\PPsi-4\beta}. 
\ee
If $\PPsi$ does not vanish  then \eqref{JJJ} can be resolved with respect to 
 the Hubble parameter, 
\be            \label{HH}
H=-\frac16\,\alpha\PPsi^3 +\frac{\beta}{3\alpha\PPsi}-\frac{C}{3\alpha\PPsi^2 a^3}\,.
\ee
Injecting this to  \eqref{Fr} yields the 
algebraic relation between $\PPsi$ and $a$, 
\be             \label{alg}
\frac{1}{24\alpha^2\PPsi^2}\left(3\alpha^2\PPsi^4-2\beta \right)\left(\alpha^2\PPsi^4-4\beta \right)
+\frac{5\alpha^2\PPsi^4-4\beta  }{6\alpha^2\PPsi^3 }\,\frac{C}{a^3}+\frac{C^2}{3\alpha^2\PPsi^4\,a^6}=\Lambda\,, 
\ee
while injecting $H$ to \eqref{FF} determines the derivative of $\PPsi$, 
\be          \label{dif}
\dot{\PPsi}=\frac{3C\PPsi(\alpha\PPsi^3-4H) }{8C-(9 \alpha^2\,\PPsi^5+4\beta\PPsi)a^3}\,.
\ee
Eqs.\eqref{HH},\eqref{alg},\eqref{dif} are invariant under 
\be               \label{symm0}
\PPsi\to -\PPsi,~~~~\alpha\to -\alpha,~~~~C\to-C;~~~~a\to a,~~~~\beta\to\beta,~~~~~H\to H, 
\ee
which provides the one-to-one correspondence between solutions of two theories which differ by 
the sign of $\alpha$. Therefore, one can assume without loss of generality that $\alpha>0$. 
{The equations are also invariant under the time reversal  $t\to -t$, which changes the sign of 
the first derivatives and of the current,  but not of the second derivatives, hence 
\be               \label{symm}
\PPsi\to -\PPsi,~~~~H\to -H,~~~~C\to-C;~~~~a\to a,~~~\dot{\PPsi}\to \dot{\PPsi},~~~~\dot{H}\to \dot{H}. 
\ee
This swaps the expanding solutions and contracting solutions. }

{
It follows  from \eqref{HH} that  if $\PPsi$ approaches zero then either the Hubble rate $H$ should  diverge,
or, if it remains finite, then the scale factor should $a$ diverge. This corresponds either 
to the initial singularity or to future infinity. 
Therefore, between these two extremities  $\PPsi$ cannot vanish and  should be sign definite, 
either everywhere positive or everywhere negative. 
}

{
One can absorb the parameters $\alpha$ and $\beta$ by expressing 
$a,\PPsi,H,\Lambda$ in terms of dimensionless\footnote{{Assuming the spacetime coordinates to have the dimension of length, $x^\mu\sim l$,  our normalisation of the action \eqref{action} implies that
$\beta,\phi,a,G_3,G_4$ are dimensionless while $\alpha^{-1}\sim X\sim K\sim \Lambda\sim l^{-2}$.}}  quantities  $x,y, h,\lambda$ via 
\be                   \label{e}
\frac{C}{a^3}=\pm \sqrt{\bm{|\beta|}}H_0\, \sqrt{x}\,y\,,~~
\PPsi=\pm H_0\,\frac{\sqrt{x}}{\sqrt{|\bm\beta|}}\, ,~~
H=\pm \frac16\,H_0\, h\,,~~
\Lambda=\frac{1}{24}\,H_0^2\,\lambda\,,
\ee
where the Hubble scale  is determined by the length scale $\sqrt{\alpha}$, 
\be                   \label{H0}
H_0=\frac{{\bm{|\beta|}}^{3/4}}{\sqrt{\alpha}}\,.
\ee
Here 
${\bm\beta}=\beta$ if $\beta\neq 0$, while if $\beta=0$  then  ${\bm\beta}$ is an {arbitrary} dimensionless parameter. 
The variable $x$ in \eqref{e} must be non-negative while $y,h,\lambda$ can be positive or negative. 
}

Injecting \eqref{e} to \eqref{KK}--\eqref{dif} yields, 
\be                   \label{ee21}
a=\left(\frac{|C| }{\sqrt{|\bm\beta|} H_0} \right)^{1/3} {\rm a} ~~~~~~~~~~~~\mbox{with}~~~~
{\rm a}=\left(\pm\frac{C}{|C|}\frac{1}{\sqrt{x}\,y}\right)^{1/3},
\ee
where the sign of $C$ should be chosen such that $\pm C/y(x)>0$, hence 
 different values of $C$ 
correspond to different solutions whose scale factors are  ``homothetic" to each other. 
\bbl{As a result,  if $C\neq 0$ then one can assume without loss of generality that either 
$C=1$ or $C=-1$, depending on sign of $y$. }
One obtains also 
\be              \label{e2}
h=\frac{2\,(\epsilon-y)-x^2}{\sqrt{x}},
\ee
while Eq.\eqref{alg} reduces to 
\be               \label{e1}
8\,y^2+(20\,x^2-16\,\epsilon)\,y+(x^2-4\,\epsilon)(3\,x^2-2\,\epsilon)=\lambda\, x\,,
\ee
with 
$$
\epsilon=
\left\{
   \begin{aligned}
    \beta/|\beta|=\pm 1 ~~~~\mbox{if}~~~~\beta\neq 0,\\
   0 ~~~~\mbox{if}~~~~\beta=0.\\
   \end{aligned}
\right. 
$$
Eq.\eqref{dif}  yields 
\be                   \label{p}
\dot{x}=2H_0\,p~~~~~\mbox{with}~~~~~
p=-\frac{5\,x^2+4\,y-4\epsilon}{9\,x^2-8\,y+4\epsilon}\,\sqrt{x}\,y\,.
\ee
The kinetic term and the sound speed in \eqref{KK} become 
\be              \label{e3}
\K&=&\frac{9\,x^2(9\,x^2-8\,y+4\epsilon) }{2\,(5\,x^2+4\,y-4\epsilon)^2}, ~~~~~~\nn \\
c_s^2&=&\frac{32\,y\,(y-7\,x^2)+(9\,x^2+4\epsilon)(4\epsilon-5x^2)}{3\,(9\,x^2-8\,y+4\epsilon)^2}.  
\ee
Eqs.\eqref{e2}--\eqref{e3} determine the solutions and their stability. 

\subsection{Currentless solutions}
\begin{figure}[h]
\hbox to \linewidth{ \hss

	
				\resizebox{8cm}{6cm}{\includegraphics{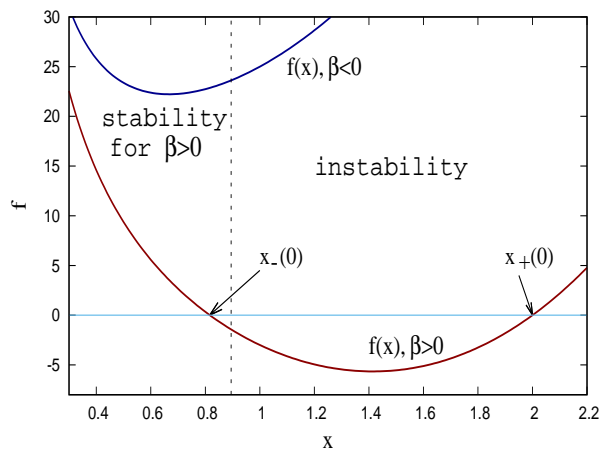}}

	
\hspace{1mm}
\hss}
\caption{
The graphical representation   of $f(x)$ in  \eqref{BOT} for $\epsilon=\beta/|\beta|=\pm 1$. 
}
 \label{Fig1}
\end{figure}

Let us first consider solutions with  vanishing scalar charge, $C=0$, in which case, according to \eqref{JJJ},
the current is zero. 
If $C=0$ then, according to \eqref{e},  one has either $x=\PPsi=0$ hence the system is in vacuum, 
or $y=0$ and then Eq.\eqref{e1}
reduces to 
\be                 \label{BOT}
f(x)\equiv \frac{ (x^2-4\,\epsilon)(3\,x^2-2\,\epsilon) }{x}=\lambda\,,
\ee
hence  $x=x(\lambda)$ is constant, $\PPsi$ and $H$ are constant as well, and
the geometry is de~Sitter. If $\epsilon= 0$ then $x(\lambda)=(\lambda/3)^{1/3}$. 
If $\epsilon =\pm 1$ then 
$f(x)$ diverges for $x\to 0,\infty$ and has a minimum in between,   hence  for $\lambda$ exceeding   
some minimal value 
there are two different solutions of \eqref{BOT}, $x=x_{+}(\lambda)$ and $x=x_{-}(\lambda)$ (see Fig.\ref{Fig1}). 

Let us assume first that $\beta>0$ hence $\epsilon=1$. Then for $\lambda=0$, 
for example,  one finds two solutions of \eqref{BOT} with the following 
properties:
\be
x_{+}(0)&=&2:~~~~~~~~h=-\sqrt{2},~~~~\K=\frac{45}{16},~~~~~c_s^2=-\frac{2}{15}; \\
x_{-}(0)&=&\sqrt{\frac{2}{3}}:~~~~h=1.476,~~~~\K=\frac{135}{2},~~~~~c_s^2=\frac{1}{45}. 
\ee
Notice that $c_s^2<0$ for the first of these solutions hence it is  unstable. 

If $\lambda$ decreases to the negative region 
then the two values $x=x_{+}(\lambda)$ and $x=x_{-}(\lambda)$ approach each other and merge 
for  $\lambda=-4\sqrt{2}$, in which case 
\be                   \label{xxx}
x_{+}(-4\sqrt{2})=x_{-}(-4\sqrt{2})=\sqrt{2},~~~~~~h=0,~~~~~~\K=\frac{11}{2}, ~~~~c_s^2=-\frac{1}{11}.
\ee
The Hubble rate vanishes for this solution and 
the geometry is flat, even though  $\PPsi\neq 0$. This solution exists only for 
$\lambda=-4\sqrt{2}$  but it is unstable  since $c_s^2<0$. 
This does not mean that flat space is always unstable in the theory, because the flat space solution can also be 
obtained in a different way: by setting $\lambda=0$ and $\PPsi=0$, in which case it is stable (as was mentioned above,
the formulas 
\eqref{e3} do not apply if $\PPsi=0$). 

Let us determine the stability region. If $y=0$ then $\K$ and $c_s^2$ defined by  \eqref{e3} 
reduce to 
\be
\K=\frac{9x^2(9x^2+4\epsilon)}{2(4\epsilon-5x^2)^2},~~~~
c_s^2=\frac{4\epsilon-5x^2}{3(9x^2+4\epsilon)}~~~~\Rightarrow~~~~\K c_s^2=\frac{3x^2}{2(4\epsilon-5x^2)}.
\ee
It follows that $\K c_s^2<0$ if $\epsilon=0,-1$ hence all solutions with $\beta\leq 0$ show either 
ghost or gradient instability.
If $\epsilon=1$ then $\K$ is always positive while $c_s^2$ will be non-negative  if 
$4-5x^2>0$, hence if  (see Fig.\ref{Fig1})
\be
~~~~~~~x\leq \frac{2}{\sqrt{5}}. 
\ee
Solutions with $x=x_{+}(\lambda)$ always violate this condition hence they are all unstable. 
Solutions with $x=x_{-}(\lambda)$ fulfill this condition if 
\be              \label{lll}
\lambda\geq -\frac{16}{5\sqrt{5}}=-1.43.
\ee
To recapitulate, the currentless solutions are characterised by a constant 
value of the scalar field gradient and by a constant Hubble rate; their geometry is 
de Sitter. For $\beta>0$ they exist if only 
$\lambda\geq -4\sqrt{2}=-5.65$  and they are stable for $\lambda\geq -1.43$. 
All of  such  solutions  for $\beta=0$ or $\beta<0$  are unstable
(we shall later  see that if the current does not vanish then stable solutions exist
for any $\beta$).

\subsection{Solutions with a non-zero current \label{curr}}
If $C\neq 0$ then the current is  $J^0=C/a^3\propto y$ 
hence the amplitude 
$y$ defined in \eqref{e} does not vanish. 
However, since $J^0\to 0$  for $a\to\infty$, it follows  that $y$ approaches zero at late times 
and the solutions then 
approach the described above configurations with constant $x$ and  de Sitter geometry. 
It follows from the above analysis that if $y$ approaches zero then $x$ must approach  either $x_{+}(\lambda)$,
in which case the product $\K c_s^2$ becomes negative and the solution becomes unstable, 
or $x$ approaches $x_{-}(\lambda)$ and then the solution is  stable if  $\lambda\geq -\frac{16}{5\sqrt{5}}$.

For $y\neq 0$ Eq.\eqref{e1} can be resolved 
yielding two different solutions, $y=y_{+}(x)$ or $y=y_{-}(x)$. 
From now on and till the end of the next sub-section we set $\epsilon=\beta/|\beta|=1$, then 
\be                   \label{yy0}
y_\pm(x)=1-\frac54\,x^2\pm\frac14\sqrt{19\, x^4-12\,x^2+2\,\lambda\,x}. 
\ee
These functions are defined only in the region where $19\, x^4-12\,x^2+2\,\lambda\,x\geq 0$. 
This region must contain a zero of $y(x)$ since we want the solution to approach for $a\to\infty$  
one of the de Sitter backgrounds described above. 
If $\lambda\geq -\frac{16}{5\sqrt{5}}$ then $y_{+}(x)$ vanishes at $x=x_{+}(\lambda)$
which point is known to be unstable, 
whereas   $y_{-}(x)$ vanishes at $x=x_{-}(\lambda)$, and we know that this point is stable. 
Therefore, we choose 
\be           \label{yy}
y(x)=y_{-}(x) 
\ee
assuming that $x\to x_{-}(\lambda)$ and hence $y\to 0$ for $a\to\infty$.  
For finite values of the universe size, when $a<\infty$, one chooses
$x\geq x_{-}(\lambda)$,  
in which case one has  $y(x)<0$. 
According to \eqref{ee21}, the  scale factor is proportional to 
\be                   \label{ee}
{\rm a}(x)=\left(\pm\frac{C}{|C|}\frac{1}{\sqrt{x}\,y(x)}\right)^{1/3},
\ee
where the sign of $C$ should be chosen such that $\pm C/y(x)>0$. 
This implies that  a$(x)\to\infty$ for $x\to x_{-}$ and 
a$(x)<\infty$ for $x> x_{-}$. 
Injecting 
\eqref{yy} to \eqref{e2} 
yields the Hubble parameter, 
\be              \label{e2a}
h(x)=\frac{2-x^2-2\,y(x)}{\sqrt{x}},
\ee
and similarly injecting to \eqref{e3} yields  $\K(x)$ and $c_s^2(x)$. 

\begin{figure}[h]
\hbox to \linewidth{ \hss

	\resizebox{7.5cm}{6.5cm}{\includegraphics{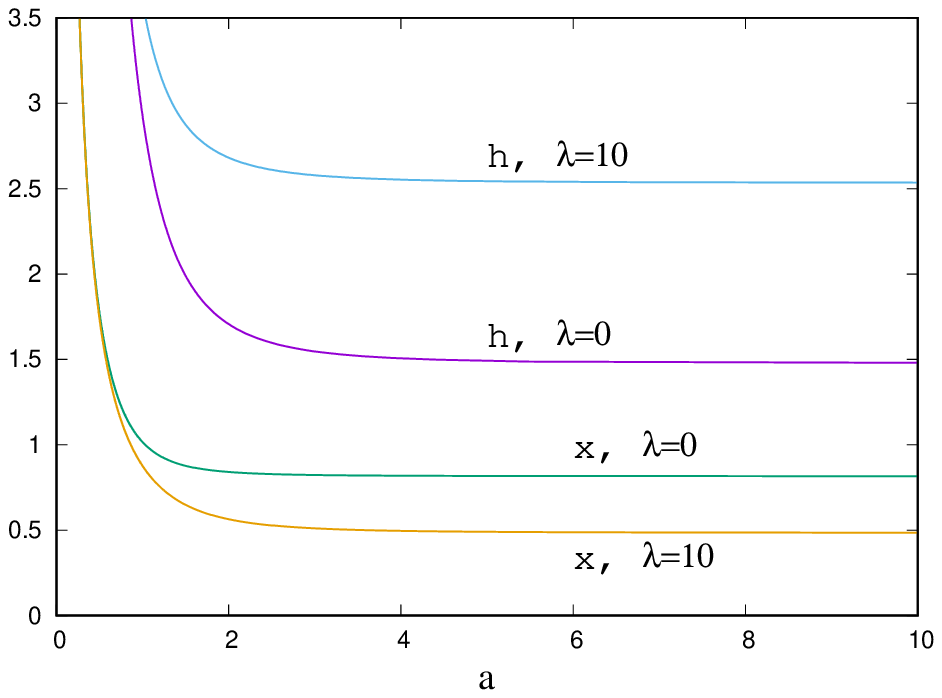}}
	\hspace{5mm}
	\resizebox{7.5cm}{6.5cm}{\includegraphics{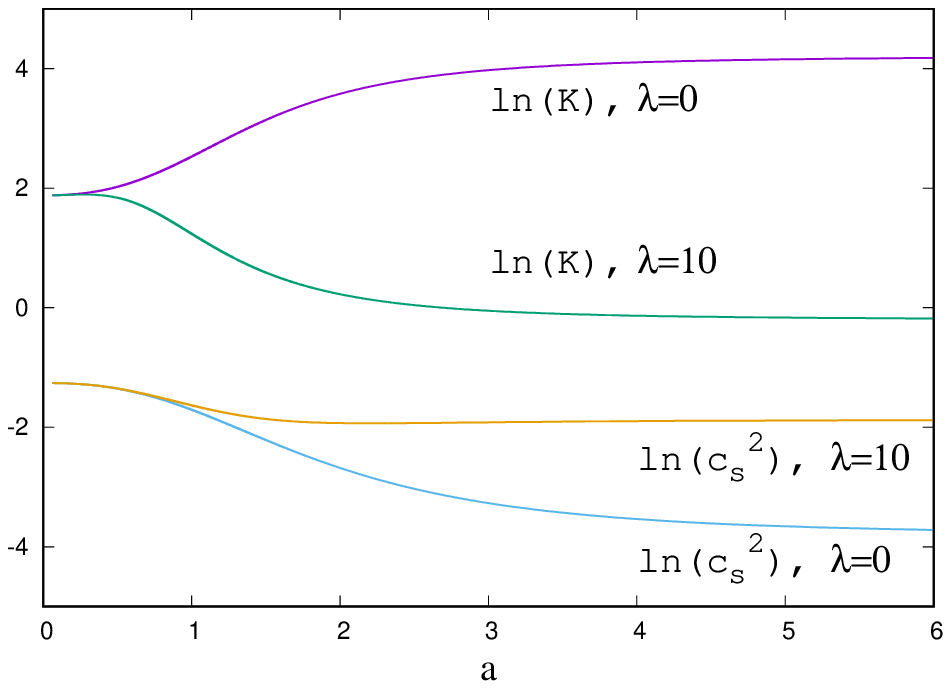}}	
		
\hspace{1mm}
\hss}
\caption{
The solution expressed by \eqref{yy},\eqref{ee},\eqref{e2a} for {$C<0$} and $\epsilon=1$. 
}
 \label{Fig2}
\end{figure}

As a result, Eqs.\eqref{yy},\eqref{ee},\eqref{e2a} provide the solution in the parametric form, 
with $x$ being the parameter.  Inverting a$(x)$ in \eqref{ee} to obtain $x=x({\rm a})$, the solution 
can be expressed in terms of the scale factor as  shown in Fig.\ref{Fig2}. 

One can see that, as the scale factor a varies from zero to infinity, the gradient squared of the 
scalar field, $x\sim\PPsi^2\sim X$, decreases from infinity to the asymptotic value $x_{-}(\lambda)$. 
The Hubble function  $h$ decreases from infinity to the constant asymptotic value defined by 
\eqref{e2} with $x=x_{-}(\lambda)$ and $y=0$.  The important point   is that the kinetic term $\K$ 
and the sound speed $c_s^2$ remain positive for all values of a, as seen in Fig.\ref{Fig2}, 
hence the solutions are always stable. 

To determine the behaviour near the initial singularity, we notice that when $x$ is large and a is small,  
then one has from \eqref{yy},\eqref{ee},\eqref{e2a} 
\be
y(x)\propto x^2,~~~~~~{\rm a}(x)\propto x^{-5/6},~~~~~~h(x)\propto x^{3/2},
\ee
hence 
\be                        \label{ww}
h^2\sim a^{-18/5}\equiv a^{-3(1+w)}. 
\ee
As a result, the system behaves  as a perfect fluid with the effective equation of state 
$w=1/5$, which is somewhere in between the dust $(w=0)$ and radiation $(w=1/3)$. 

To recapitulate, the system admits cosmological solutions with a non-zero scalar current.
Close to the initial singularity, the squared gradient 
of the scalar field is $X\propto a^{-6/5}$,  which  mimics   a perfect fluid with the equation of state $w=1/5$. 
As the size of the universe  grows, $X$ and the 
Hubble rate approach constant values. 
These solutions are stable. 

It is worth noting that these solutions can describe both the expansion and contraction of the universe, 
according to the choice of sign in  Eq.\eqref{e}. Choosing the plus sign yields $H>0$, hence the expansion, 
in which case one should choose $C<0$ since $y<0$. Choosing the minus sign gives the contraction with $H<0$,
 and then one should choose $C>0$. The two cases are related by the symmetry \eqref{symm}. 

\subsection{More general solutions}

These are defined by the algebraic curve 
 $y(x)$ subject to  \eqref{e1}. To study this curve, we plot together the functions
$y_{+}(x)$ and $y_{-}(x)$ defined by \eqref{yy0}, which allows us to distinguish different solution types.  
Depending on value of $\lambda$, these solutions  can be classified as follows.

\subsubsection{$\lambda<-4\sqrt{2}=-5.65$} 

\begin{figure}[h]
\hbox to \linewidth{ \hss

	\resizebox{7.5cm}{6.5cm}{\includegraphics{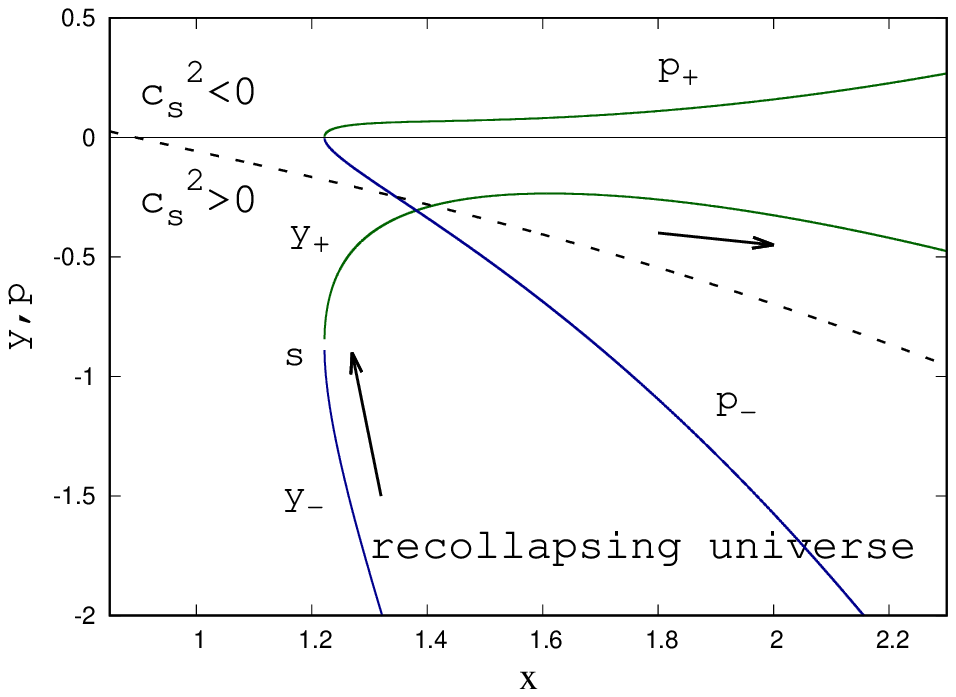}}
	\hspace{5mm}
	\resizebox{7.5cm}{6.5cm}{\includegraphics{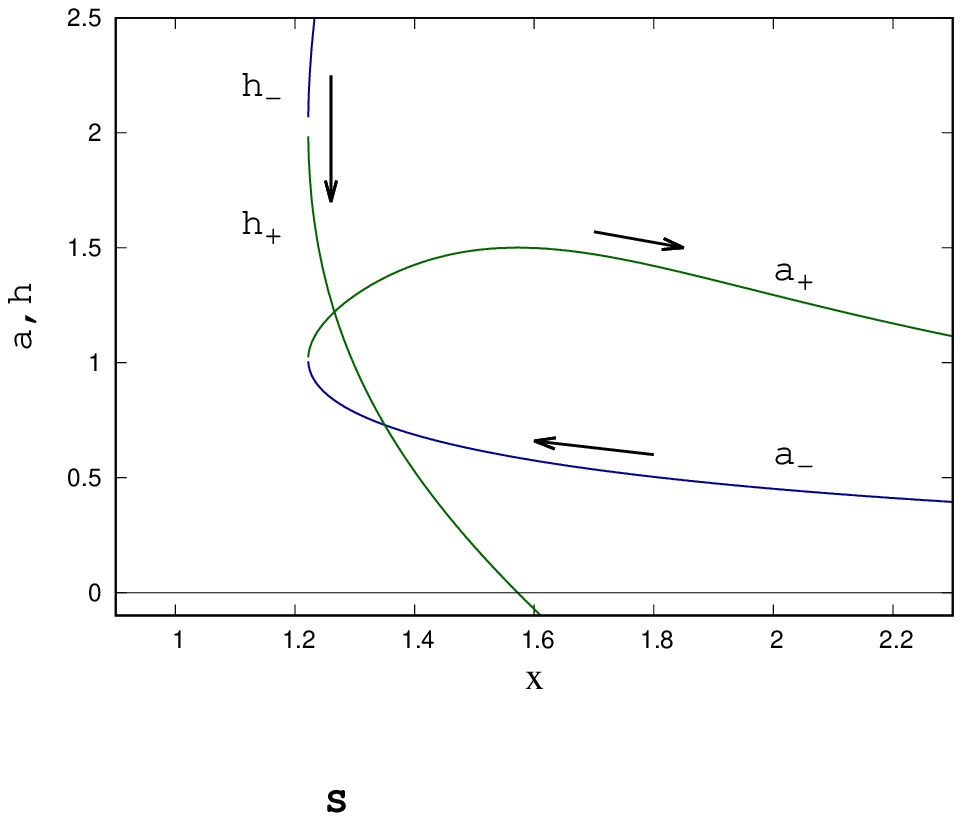}}	
		
\hspace{1mm}
\hss}
\caption{
The functions $y(x)=y_\pm(x)$ and  $p(x)$, $a(x)$, $h(x)$ defined by \eqref{yy0},  \eqref{p}, \eqref{ee}, \eqref{e2a} 
for $\lambda=-10$ and $\epsilon=1$. \bbl{The dashed line marks the border where the sound speed $c_s^2$ 
defined by  \eqref{e3}
vanishes, hence the part of the solution to the right from this line shows the gradient instability.  }
}
 \label{Fig3}
\end{figure}

\underline{Type I.} An example of such solutions is shown in Fig.\ref{Fig3}
for $\lambda=-10$.  Both $y_{+}(x)$ and $y_{-}(x)$ are everywhere negative and defined only in the region 
$x\geq x_{\rm min}(\lambda)$ where the argument of the square root in \eqref{yy0} is positive.
In the $x\to x_{\rm min}(\lambda)$ limit the square root vanishes and the  $y_{+}(x)$ and $y_{-}(x)$ 
branches merge at the point $s$ marked on the left panel in Fig.\ref{Fig3}. Nothing special happens 
at this point:  the solution simply passes from the lower $y_{-}(x)$ branch to the upper
 $y_{+}(x)$ branch in the direction indicated by the arrow in Fig.\ref{Fig3}. 
 The direction is determined by the fact that at the lower branch the derivative $\dot{x}\propto p_{-}<0$, 
 as shown in Fig.\ref{Fig3}, hence 
 $x$ decreases towards the minimal value $x_{\rm min}(\lambda)$, while at the upper branch the 
 derivative $\dot{x}\propto p_{+}>0$ and $x$ increases. 
 
 The scale factor ${\rm a}(x)$ obtained from \eqref{ee} increases along the lower branch and the 
 corresponding Hubble parameter is positive, $h_{-}>0$, as shown on the right panel in Fig.\ref{Fig3}. 
 After passing to the upper branch,  the scale factor first continues to increase up to a maximal value,
 then the Hubble parameter $h_{+}$ changes sign and the universe starts shrinking. 
 
Therefore, the universe starts from zero size at $x=\infty$ and  $y=y_{-}=-\infty$,
 then it expands first along the $y_{-}$ branch and next along the $y_{+}$ branch, then the scale 
 factor reaches a maximal finite value, after which the universe shrinks back to  zero size 
 along the $y_{+}$ branch. The sound speed $c_{s}^2$ becomes negative at the $y_{+}$ branch, 
 hence the solution is unstable. 
 
The solution remains qualitatively the same for any $\lambda<-4\sqrt{2}=-5.65$, when both $y_{+}$ 
and $y_{-}$ remain negative, but the maximal value of $y_{+}$ approaches zero from below 
when $\lambda$ increases. 

\subsubsection{$\lambda=-4\sqrt{2}$ } 

\begin{figure}[h]
\hbox to \linewidth{ \hss

	\resizebox{7.5cm}{6.5cm}{\includegraphics{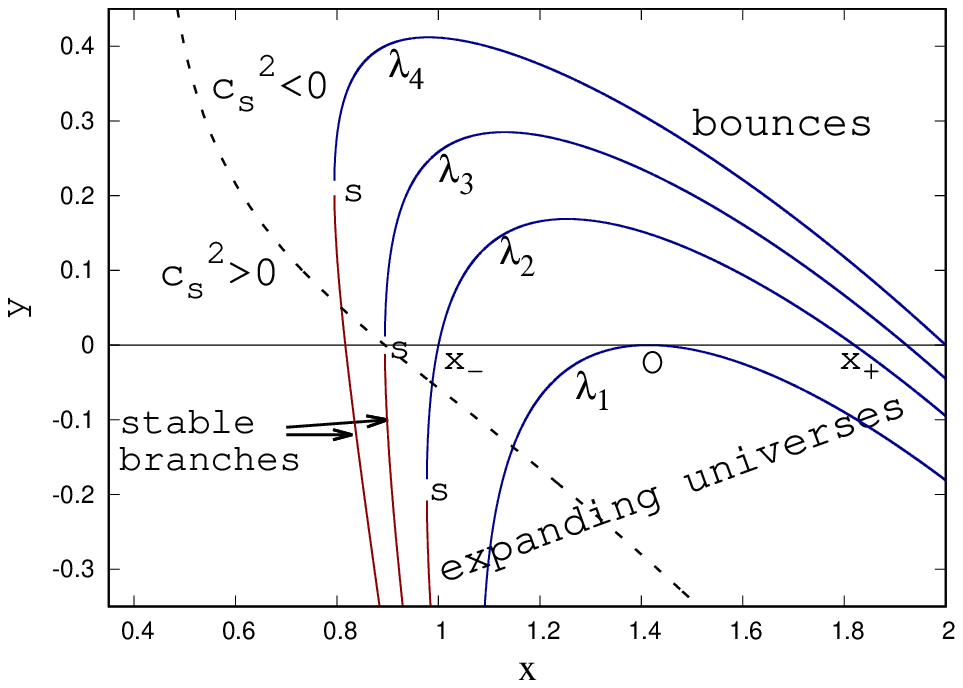}}
	\hspace{5mm}
	\resizebox{7.5cm}{6.5cm}{\includegraphics{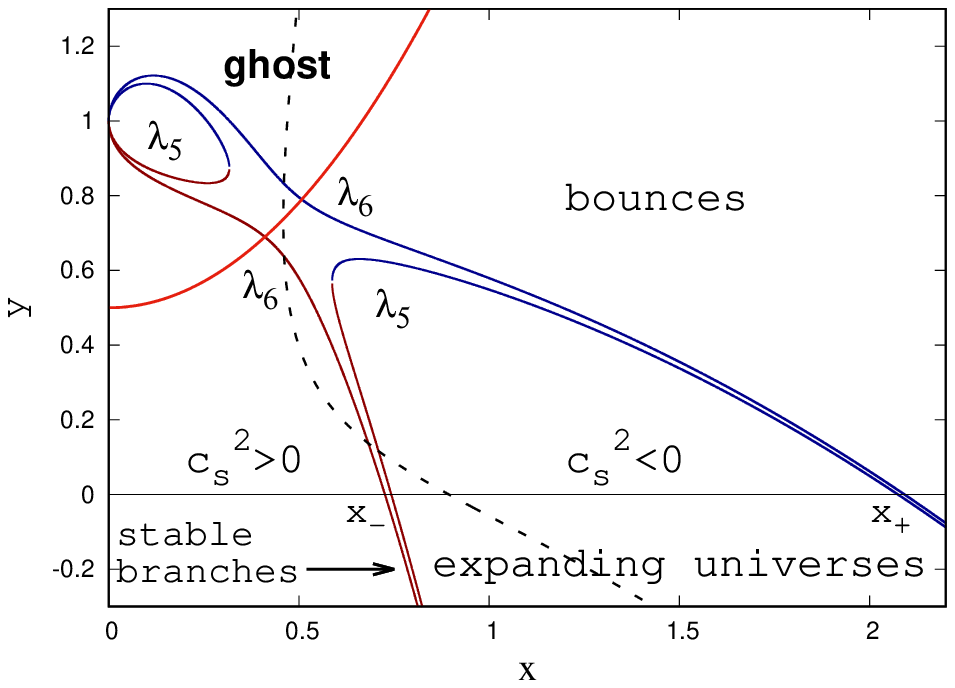}}	
		
\hspace{1mm}
\hss}
\caption{
The curve $y(x)=y_{-}(x)\cup y_{+}(x)$ defined by \eqref{yy0} for $\lambda$ equal to
$\lambda_1=-4\sqrt{2}$, $\lambda_2=-3$, $\lambda_3=-16/(5\sqrt{5})$, $\lambda_4=0$, 
$\lambda_5=1.6<8/\sqrt{19}$ and $\lambda_6=2>8/\sqrt{19}$. \bbl{The $y>0$ parts of the curves correspond to bounces 
while the $y<0$ parts describe expanding universes. 
All bounce curves  intersect the $c_s^2=0$ line and show the gradient 
instability, while those in the right panel  intersect in addition the red line
 where K defined by \eqref{e3} vanishes and hence they enter the ghost region. All expanding universes 
 corresponding to the right parts of the $y<0$ curves show gradient instability, but the left parts of the curves 
 stay away from the tachyon region  and describe stable cosmologies if $\lambda\geq -16/(5\sqrt{5})$. 
}
}
 \label{Fig4}
\end{figure}

\underline{Type II.} The curve $y(x)=y_{-}(x)\cup y_{+}(x)$  remains 
qualitatively the same as before but the $y_{+}(x)$ branch touches zero from below at the 
point O indicated on the left panel in Fig.\ref{Fig4}. The position of this point is described 
by Eq.\eqref{xxx} above.  Since $y$ vanishes at this point, the universe size \eqref{ee} becomes infinite. 
Therefore, there are actually two different solutions in this case. The part of the $\lambda_1$-curve 
on the left panel in Fig.\ref{Fig4} which is on the left from the point O describes the universe expanding 
from zero size to infinity. The part of the curve on the right from O describes the universe 
shrinking from infinite size to zero. 
Since at the point O the sound speed squared becomes negative (see Eq.\eqref{xxx}), 
both of these solutions show a gradient instability. 

\subsubsection{$-4\sqrt{2}<\lambda \leq 0$}
The $y(x)=y_{-}(x)\cup y_{+}(x)$ curve remains 
qualitatively the same as before but shifts upwards and crosses zero twice at $x=x_{+}$ and 
$x=x_{-}$ with $x_{\pm}(\lambda)$ defined by Eq.\eqref{BOT}. This is the case for the 
$\lambda_2$, $\lambda_3$, and $\lambda_4$ curves
on the left panel in Fig.\ref{Fig4}. 
Since $y(x_{\pm})=0$, it follows that 
$a_\pm(x_\pm)=\infty$ (see \eqref{ee}),  hence the single curve $y(x)$ 
determines three different solutions of types III, IV, V described below. 

\underline{Type III.} This solution corresponds to the left part of the $y(x)$ curve 
where $y(x)<0$ (the $\lambda_2$, $\lambda_3$, and $\lambda_4$ curves
in Fig.\ref{Fig4}). 
This determines the universe 
expanding from zero to infinity. This solution can be stable or unstable, depending on the 
position of the point $s$ of the merging of the  $y_{+}(x)$ and $y_{-}(x)$. If the merging point is below 
the $x$-axis (as for the $\lambda_{2}$-curve in Fig.\ref{Fig4}) then the solution is unstable. 
The solution becomes stable for $\lambda=-16/(5\sqrt{5})=-1.43$ 
when the merging point $s$ is at the $x$-axis 
(the $\lambda_{3}$-curve in Fig.\ref{Fig4}), 
and it remains stable when $s$  moves further up
 (the $\lambda_{4}$-curve in Fig.\ref{Fig4}). 
The profiles of the solution in the latter two cases are similar to those shown in Fig.\ref{Fig2}. 

\underline{Type IV.} This solution corresponds to the part of the $y(x)$ curve
interpolating between $x=x_{-}$ and $x=x_{+}$ (the $\lambda_2$, $\lambda_3$, and $\lambda_4$ curves
in Fig.\ref{Fig4}). This describes a bounce -- the universe 
shrinking to a finite  size and then expanding back to infinity. This solution is always 
unstable since it contains the point $y=0$, $x=x_{+}(\lambda)$ which is known to be unstable. 

\underline{Type V.} This solution corresponds to the right part of the $y(x)$ curve where $y<0$,
(the $\lambda_2$, $\lambda_3$, and $\lambda_4$ curves
in Fig.\ref{Fig4}). This also corresponds to the universe expanding from zero to infinity 
(or contracting, depending on the sign choice in \eqref{e}), and it is always unstable 
because  it contains the unstable point $y=0$, $x=x_{+}(\lambda)$. 

\subsubsection{$0<\lambda < 8/\sqrt{19}=1.83$}
The $y(x)=y_{-}(x)\cup y_{+}(x)$ curve moves  further upwards and develops   a disjoint part -- 
a small loop, as illustrated by the $\lambda_5$-curve shown 
on the right panel in Fig.\ref{Fig4}. The curve therefore splits into two disconnected subsets -- 
the compact part (the loop) and the non-compact part. 
The non-compact part corresponds to three different solutions of 
 Types III--V described above; Type III solution always being stable. 
 The compact part corresponds to a new solution type 
with the following properties. 

{
\underline{Type VI.} The small loop shown in  Fig.\ref{Fig4} touches the vertical axis at the point $(x,y)=(0,1)$
where the universe size $a$ diverges. 
In the vicinity of this point one has 
\be                    \label{rip}
a\sim x^{-1/6},~~~
\frac{\dot{a}}{a}\sim h_\pm=\mp \sqrt{\frac{\lambda}{2}}+{\cal O}(x),~~~~
~~~\Rightarrow ~~~~a\sim e^{\mp {H}t},
\ee
where ${H}=\sqrt{\lambda}H_0/(6\sqrt{2})$. 
The evolution along the loop corresponds to the universe starting from an infinite size $a\sim e^{- {H}t}$
in the past,  then shrinking to a finite size, bouncing back  and expanding again to 
an infinite size $a\sim e^{+ {H}t}$. 
These  solutions show ghost. 
}

\subsubsection{$8/\sqrt{19}<\lambda$}
If $\lambda$ exceeds  the value $8/\sqrt{19}$ then 
the two disjoint pars of the $y(x)$ curve interconnect to  form one connected manifold, as illustrated by the 
$\lambda_6$ curve in  Fig.\ref{Fig4}. This corresponds to four different solutions. 
The two parts of the curve where $y(x)\leq 0$ correspond to solutions of Types III and V;
 Type III  always being stable. The parts of the curve where $y(x)\geq 0$ correspond to two 
 different solutions of the following new type. 

 \underline{Type VII.} The two parts of the $y(x)$-curve which interpolate between 
 points $(x,y)=(0,1)$ and $(x_{-},0)$ 
 or between $(0,1)$ and $(x_{+},0)$ 
correspond to bounces -- the universe starts from 
 and ends up with an
 infinite size. 
 These solutions are unstable. 
 
 
 Summarizing, the only stable solutions in the above classification are those of Type III; 
 they exist only for $\lambda\geq -16/(5\sqrt{5})=-1.43$. They are qualitatively 
 the same as those previously described 
 in sub-section \ref{curr}.

 \subsection{Solutions with $\beta\leq 0$}
 
\begin{figure}[h]
\hbox to \linewidth{ \hss

	\resizebox{7.5cm}{6.5cm}{\includegraphics{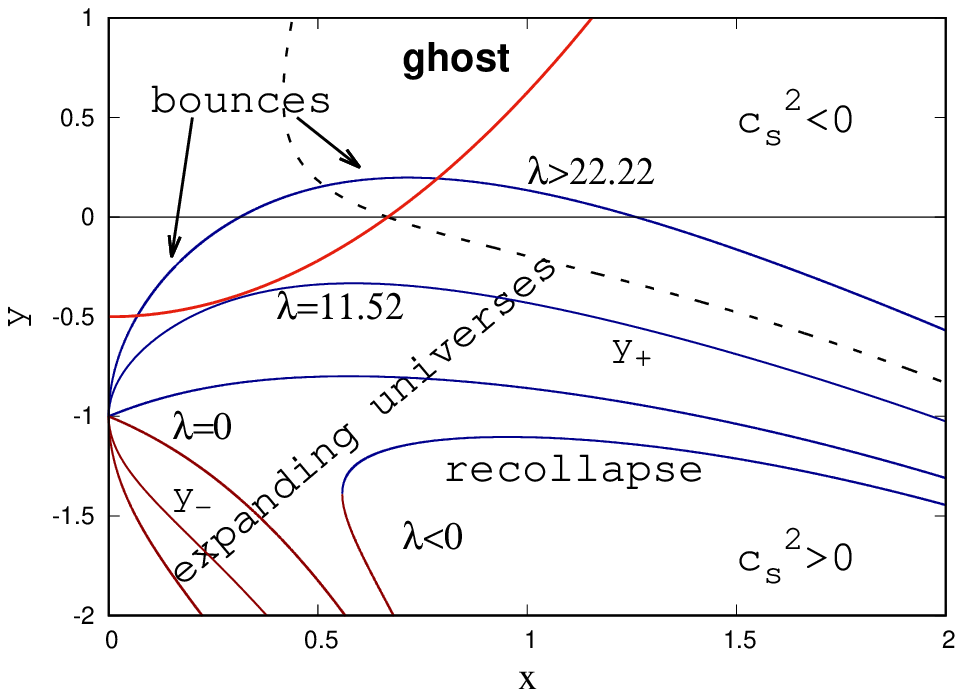}}
	\hspace{5mm}
	\resizebox{7.5cm}{6.5cm}{\includegraphics{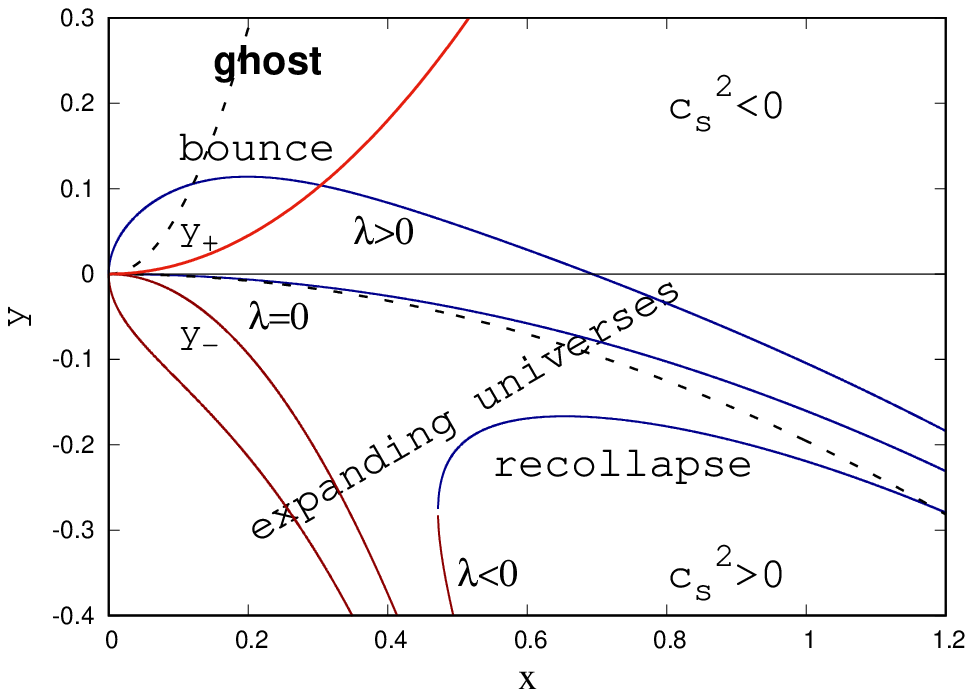}}	
		
\hspace{1mm}
\hss}
\caption{The function $y(x)=y_{+}(x)\cup y_{-}(x)$ defined by \eqref{yy0a} for 
 $\beta<0$ (left) and  $\beta=0$ (right). \bbl{The right parts of the curves (marked by $y_{+}$) always intersect the 
 $c_s^2=0$ line for large enough $x$, hence the corresponding solutions are unstable; some of them also 
 enter the  ghost region. The left parts of the curves marked by $y_{-}$ always keep away from the 
 instability regions  and describe stable expanding cosmologies if $\lambda\geq 0$. }
}
 \label{Fig5}
\end{figure}

Solving Eq.\eqref{e1} for $\epsilon=0$ or $\epsilon=-1$ yields 
\be                   \label{yy0a}
y_\pm(x)=\epsilon-\frac54\,x^2\pm\frac14\sqrt{19\, x^4-12\epsilon \,x^2+2\,\lambda\,x}, 
\ee
which should be injected to \eqref{ee}, \eqref{e2a} and to \eqref{e3} to determine the solutions. 
The behaviour of $y(x)=y_{+}(x)\cup y_{-}(x)$ is shown in Fig.\ref{Fig5}, and this time 
one finds  only two qualitatively different solution 
types. First, if $\lambda<0$ then $y(x)$ is illustrated by the $\lambda<0$ curve 
shown on the left panel in  Fig.\ref{Fig5}. This 
gives rise  to Type I solutions  described above, they   are always unstable. 

If  $\lambda\geq 0$ then both $y_{+}(x)$ and $y_{-}(x)$ curves touch  the 
vertical axis at the point  with coordinates $(0,\epsilon)$,
which corresponds to 
$a=\infty$ (see Fig.\ref{Fig5}), hence the solution splits in two.  One solution
is generated by curves $y_{-}(x)$ which emanate from $(0,\epsilon)$ downwards.
 These solutions are stable. 
 The other solutions are generated by curves $y_{+}(x)$ emanating
  from $(0,\epsilon)$ towards increasing values of $y$, all of them are unstable.

  Let us describe the stable solutions.
 If $\epsilon=-1$ and $\lambda\neq 0$ then the Hubble parameter 
 at the point $(0,\epsilon)$ becomes $h_{-} =\sqrt{\frac{\lambda}{2}}$
 (see \eqref{e2}), which corresponds to the behaviour \eqref{rip}. 
 Therefore, the $y_{-}$ curves for $\lambda\geq 0$ shown in the 
 lower left corner on the left panel in Fig.\ref{Fig5} describe  the universe starting from 
 zero size in the past and expanding in the future as  $a\sim e^{{H}t}$
 with  ${H}=\sqrt{\lambda}H_0/(6\sqrt{2})$. 
 These solutions are stable. 
 
  If $\epsilon=-1$ and $\lambda=0$ then one has for  small $x$
 \be                  \label{w4}
 h\sim \sqrt{x}, ~~~~a\sim x^{-1/6}~~~~\Rightarrow~~~h^2\sim \frac{1}{a^6}\equiv \frac{1}{a^{3(1+w)}}~~~\Rightarrow ~~
 a\sim t^{1/3}. 
 \ee
 Therefore, the $y_{-}$ curve for $\lambda=0$ in the  left 
 lower corner on the left panel in Fig.\ref{Fig5} describes the universe starting from 
 zero size in the past and entering in the future the $a\sim t^{1/3}$ regime 
 corresponding  to the $w=1$ equation of state.

 If $\epsilon=0$ but $\lambda\neq 0$ then at small $x$ one has 
  \be
  h_{-} =\sqrt{\frac{\lambda}{2}}+{\cal O}(x^{3/2}),~~~~a\sim x^{-1/6}~~~~\Rightarrow~~~~
 a\sim e^{-{H}t},
 \ee
 where ${H}=\sqrt{\lambda}H_0/(6\sqrt{2})$.  
  Therefore, the $y_{-}$ curves for $\lambda>0$ shown in the 
 lower left corner on the right panel in Fig.\ref{Fig5} correspond to the universe starting from 
 a zero size in the past and approaching asymptotically the de Sitter phase. 
The squared gradient of the scalar field $X\sim \PPsi^2\sim x$  asymptotically 
 approaches zero. These solutions are stable.

{
 If $\epsilon=\lambda=0$ then  for any $x$ one has
 \be
 y_{-}=-\frac{5+\sqrt{19}}{4}\,x^2,~~~h=\frac{3+\sqrt{19}}{2}\,x^{3/2},~~~a\sim x^{-5/6},
 \ee
  therefore at all times the universe exactly follows the $w=1/5$ equation of state
 \be                   \label{www} 
 h^2\sim a^{-18/5}\equiv a^{-3(1+w)}~~~~\Rightarrow~~~~a\sim t^{5/9}. 
 \ee
 This type of behaviour we have already seen in \eqref{ww} close to the singularity, but this time 
 it holds everywhere. 
  This solution is stable. 
 }
  
Summarizing, stable for $\beta\leq 0$  solutions exist for $\lambda\geq 0$ and 
are generated by the $y_{-}(x)$ curves residing 
in the lower left corners in the diagrams in Fig.\ref{Fig5}. They describe universes expanding from zero size 
to infinity. If $\beta<0$ then the universe approaches in the future  the de Sitter phase with the Hubble rate 
${H}=\sqrt{\lambda}H_0/(6\sqrt{2})$ if $\lambda\neq 0$, while for $\lambda=0$ it expands at late times according to the 
$w=1$ equation of state. 
For $\beta=0$ and $\lambda>0$  the universe approaches the de Sitter phase with the same Hubble rate 
${H}=\sqrt{\lambda}H_0/(6\sqrt{2})$ if $\lambda\neq 0$,
 whereas for $\beta=0$ and $\lambda=0$  it expands at all times 
according to the $w=1/5$ equation of state. 

\subsection{Solutions in the metric version of the theory}

Let us now compare studied above solutions in the Palatini version of the theory with those 
arising in the metric version of the theory. As was mentioned, the equations of the metric version 
can be obtained from \eqref{Fr}, \eqref{FF}, \eqref{Jzero}, \eqref{JJJ} by omitting  terms proportional 
to $\alpha^2$. Applying the rescaling \eqref{e} then yields the modified version of 
Eqs.\eqref{e2}, \eqref{e1}, \eqref{p}, 
\eqref{e3}. The equation for $y$ becomes 
\be                    \label{A1}
8\,y^2+(12\,x^2-16\,\epsilon)\,y+2\epsilon (4\epsilon-3x^2)=\lambda\, x\,, 
\ee
and one has 
\be                   \label{A2}
h=\frac{2(\epsilon-y)}{\sqrt{x}}, ~~~~~
p=-\frac{(3\,x^2+4\,y-4\epsilon)y}{3\,x^2-8\,y+4\epsilon}\,\sqrt{x},~~~~~
{\rm a}=\left(\pm\frac{C}{\sqrt{x}\,y}\right)^{1/3}.
\ee
\begin{figure}[h]
\hbox to \linewidth{ \hss

	\resizebox{7.5cm}{6.5cm}{\includegraphics{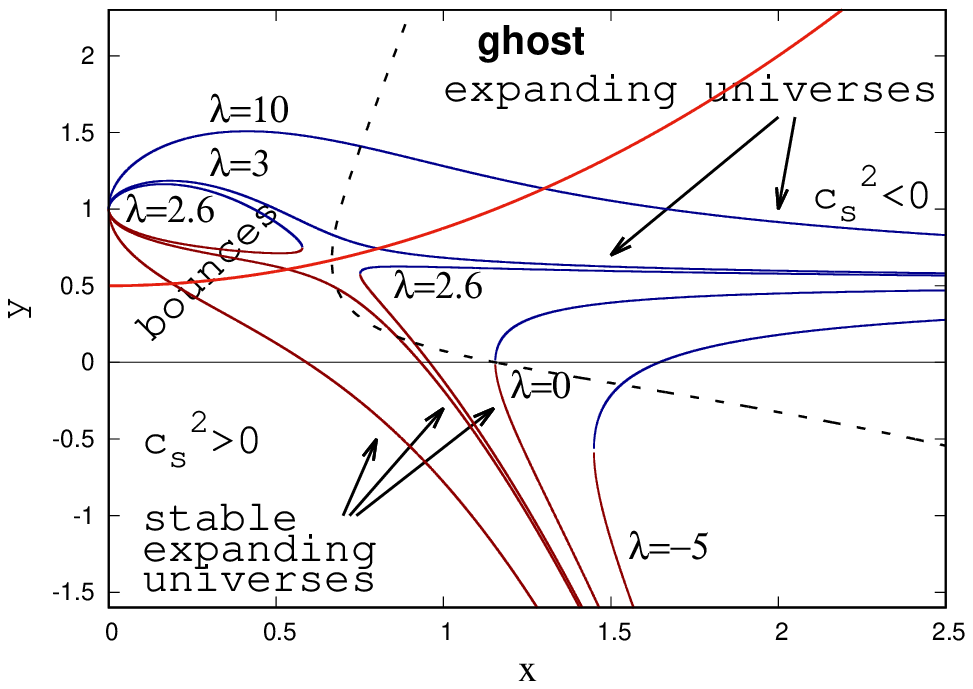}}
	\hspace{5mm}
	\resizebox{7.5cm}{6.5cm}{\includegraphics{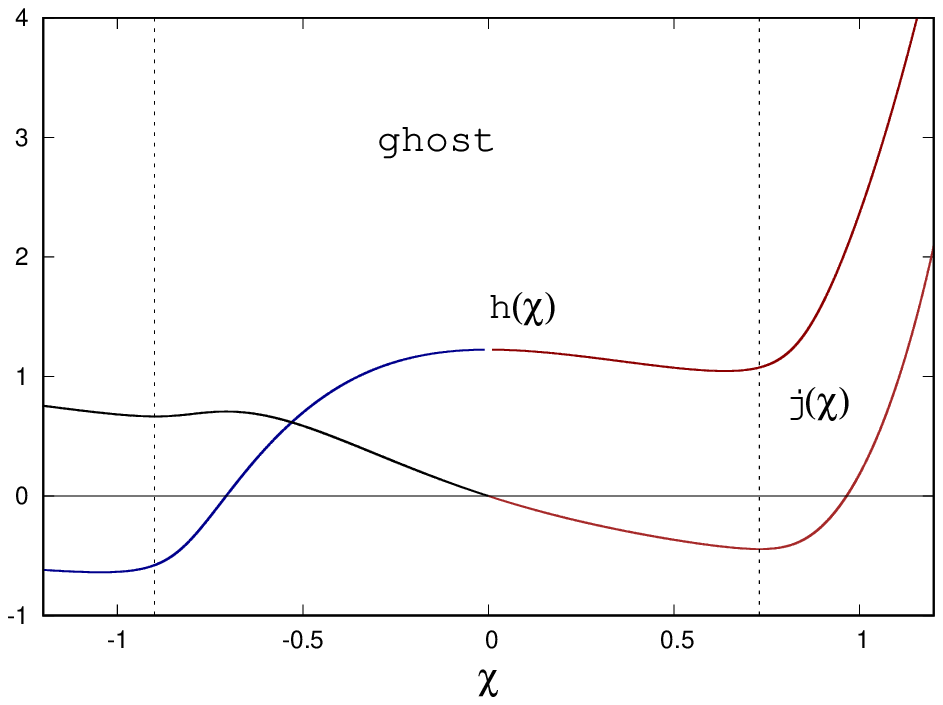}}	
		
\hspace{1mm}
\hss}
\caption{Left: solutions of \eqref{A1} for $\beta>0$ ($\epsilon=1)$. 
\bbl{If $y\to 0$ and/or $x\to 0$ then $a\to\infty$, while if $y\to\infty $ and/or $x\to\infty $ then $a\to 0$,
hence the $y(x)$ curves describe either expanding cosmologies or bounces. 
The $y\leq 0$ parts of the  curves describe stable expanding cosmologies if $\lambda\geq 0$; all other solutions are unstable. }
Right: profiles of the $\lambda=3$ solution 
against the dimensionless $\chi\sim\dot{\phi}$. 
}
 \label{Fig6}
\end{figure}
The properties of perturbations are read-off from \eqref{K},\eqref{XXX}, after replacing in these formulas 
$\kappa\to \kappa-2 XA$. 
 This yields the kinetic term and sound speed:
\be                   \label{A3}
\K=\frac{9\,x^2(3x^2-8y+4\epsilon)}{2\,(3x^2+4\,y-4\epsilon)^2},~~~~~~
c_s^2=\frac{32\,y\,(y-3x^2)-9x^4+16\epsilon^2}{3\,(3\,x^2-8\,y+4\epsilon)^2}\,.
\ee
The procedure is then the same as before: first one solves \eqref{A1} to obtain 
$y(x)=y_{+}(x)\cup y_{-}(x)$ with
\be
y_\pm(x)=\epsilon-\frac34\,x^2\pm \frac14 \sqrt{9x^4-12\epsilon x^2+2\lambda x }.
\ee
This determine algebraic curves shown in Figs.\ref{Fig6},\ref{Fig7} (in the online version of Figs.\ref{Fig4}--\ref{Fig7}
the $y_{+}(x)$ and $y_{-}(x)$ amplitudes are  shown, respectively, in dark-blue and dark-red). 
The  interpretation of these curves  is obtained by injecting $y_\pm(x)$ to \eqref{A2} and \eqref{A3}: 
for example, 
points where $y(x)$ either crosses the horizontal axis or touches the vertical axis correspond to the 
infinite size of the universe. As a result, the curves in  Figs.\ref{Fig6},\ref{Fig7}  corresponds 
either to universes expanding from zero to infinite size, or to universes expanding only up to a finite size and then shrinking, or to bounces. 
\begin{figure}[h]
\hbox to \linewidth{ \hss

	\resizebox{7.5cm}{6.5cm}{\includegraphics{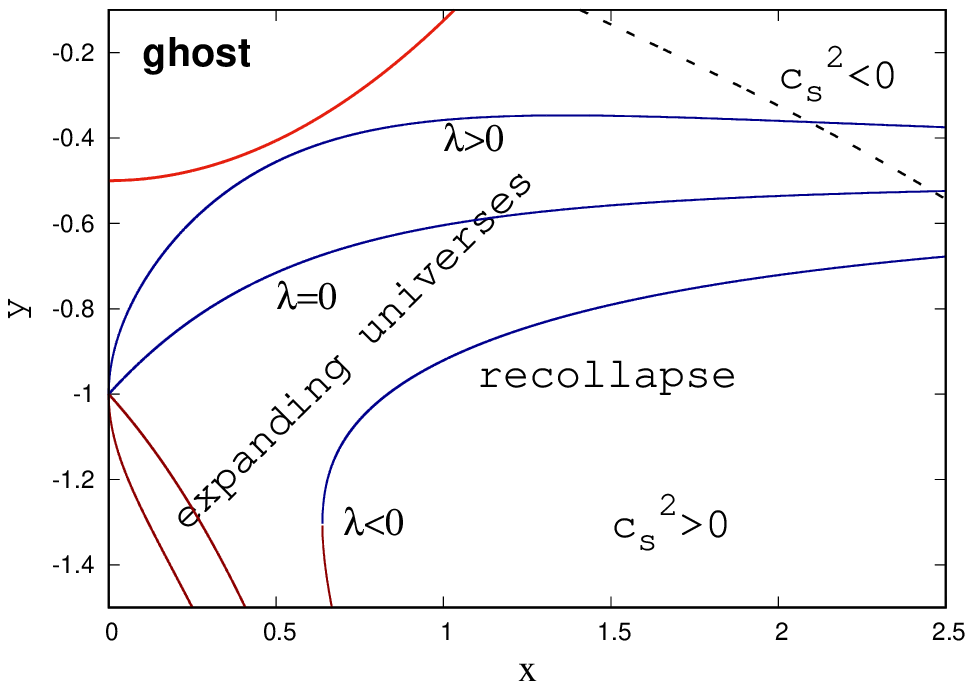}}
	\hspace{5mm}
	\resizebox{7.5cm}{6.5cm}{\includegraphics{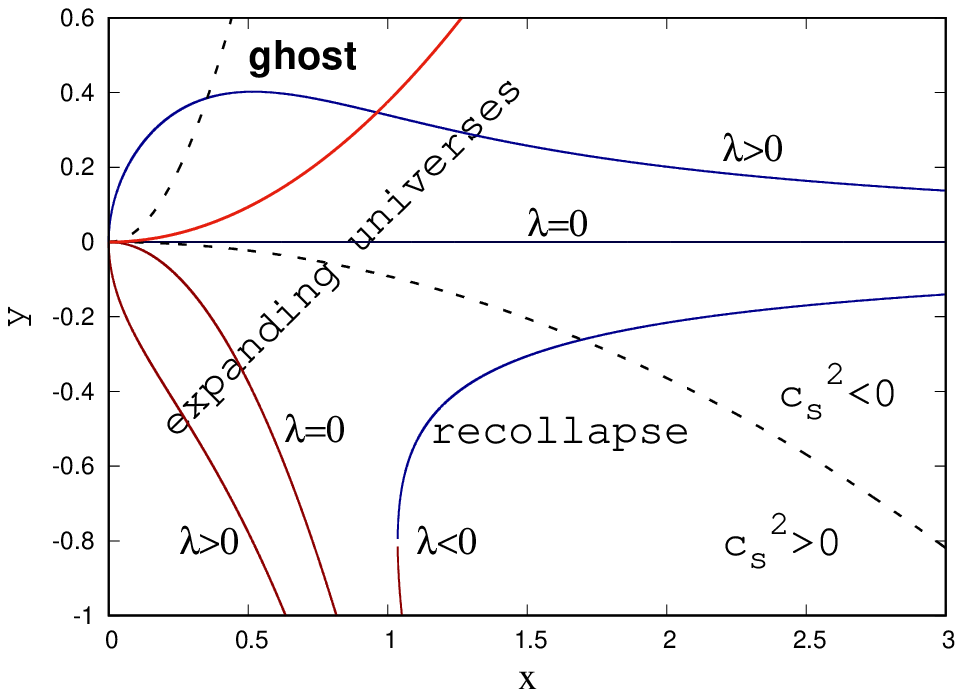}}	
		
\hspace{1mm}
\hss}
\caption{Solution of \eqref{A1} for $\beta<0$ (left) and for $\beta=0$ (right). 
\bbl{They describe either expanding or recollapsing universes. The  solution  branches 
corresponding  to the left lower parts of the curves that touch  the $y$-axis exist for $\lambda\geq 0 $
and describe stable cosmologies.}
}
 \label{Fig7}
\end{figure}

The  {\it stable} solutions are again only those generated by the parts of the $y_{-}(x)$ curves 
located under the horizontal axis 
in the left lower corner of the diagrams  in  Figs.\ref{Fig6},\ref{Fig7}.  
Such solutions exist for any $\epsilon=0,\pm 1$ 
but  only for $\lambda\geq 0$. Their profiles are qualitatively similar to those
shown in Fig.\ref{Fig2}. The overall conclusion is that, despite their surprising variety, 
solutions in  the Palatini-derived theory and those of the metric theory are qualitatively similar to each other. 

\subsection{Summary and comparison with the previously studied case}

\bbl{We presented above {\it all} homogeneous and isotropic cosmologies
 in two special shift-symmetric KGB models, one of which 
is the standard metric theory with the coefficient functions 
\be              \label{kgb1}
G_3=\alpha X,~~~~~~~K=\beta X-2\Lambda.
\ee  
The other one is its Palatini version, which, according to the discussion in Section \ref{SRel}, 
can also be viewed as the standard metric theory with 
\be             \label{kgb2}
G_3=\alpha X,~~~~~~~K=\beta X-\frac23\,\alpha^2 X^3 -2\Lambda.
\ee  
As we have seen, the classification of solutions in these two theories is rather complicated, but 
the solutions are sensitive only to $\Lambda$ 
and to the sign of $\beta$, while $\alpha$ only changes the scale. Despite some differences 
between the two theories seen in the figures above,  the properties  of the 
solutions turn out to be essentially the same in both cases and the solutions are always found to be either 
expanding or recollapsing universes, or bounces. }

\bbl{The most interesting physically are the stable expanding cosmologies. 
 They exhibit the effective equation of state $w=1/5$ near singularity 
and at late times they approach the currentless de Sitter phase with a constant Hubble rate.
If $\beta< 0$ or $\beta=0$  then the  Hubble rate is expressed in terms of $\Lambda$ in the usual way, 
$H=\sqrt{\Lambda/3}$, but for $\beta>0$  it is not directly related to $\Lambda$, 
and if $\Lambda=0$ then 
\be
H=c\times \left(\frac{\beta^3}{\alpha^2}\right)^{1/4}
\ee
where the numerical coefficient  $c=1.64\times 10^{-1}$ in model \eqref{kgb1} and $c=3.65\times 10^{-3}$
in model \eqref{kgb2}. The theories also show recollapses and bounces,  
but these solutions are always unstable. }

\bbl{
To the best of our knowledge, a similar complete  classification of KGB cosmologies  
has never been reported in the literature.  
At the same time, it was known before that the equations are completely integrable in 
the shift-symmetric case,  and  cosmologies were much studied
from a  more  physical perspective
already in the  original KGB paper \cite{Deffayet:2010qz}. 
}
 
 \bbl{
 The specific model considered in Ref.\cite{Deffayet:2010qz} corresponds to \eqref{kgb1} 
 with  $\beta>0$\footnote{\bbl{See Eq.(5.1) in \cite{Deffayet:2010qz} where one should replace $X\to -X$ since 
 the metric signature used in that paper is opposite to ours. }} 
 but with  $\Lambda$ 
 replaced by an 
  external matter energy density, $\Lambda\to \rho_{ext} \sim a^{-3(1+w_{ext})}$. 
 The case of a constant $\rho_{ext}$ was also considered there,  
and it seems that what is shown in Fig.3 in Ref.\cite{Deffayet:2010qz} corresponds to our solutions with $\epsilon=1$ 
and $\lambda=3$ on the left panel of our  Fig.\ref{Fig6}. 
The $\lambda=3$ curve shown there describes 
three different solutions: a stable cosmology, an unstable cosmology, and a 
bounce, the two latter showing the ghost. 
Now, on the right panel in  Fig.\ref{Fig6} we plotted 
profiles of these solutions against the dimensionless variable $\chi\sim \psi$ defined as $\chi=\sqrt{x}$ in the $\chi>0$ 
region and $\chi=-\sqrt{x}$ in the $\chi<0$ region. We used the symmetry \eqref{symm} to relate the
 values of the solutions for opposite signs of $\PPsi$. The dimensionless Hubble parameter 
 shown in Fig.\ref{Fig6} is defined as $h(\chi)=h_{-}$ for $\chi>0$ and $h(\chi)=-h_{+}$ for $\chi<0$,
 while the dimensionless current is $j(\chi)=-\sqrt{x}y_{-}$ for $\chi>0$ and 
 $j(\chi)=\sqrt{x}y_{+}$ for $\chi<0$. The vertical lines delimit the instability region where $K<0$. 
 The resulting diagram 
  is very similar to Fig.3 in \cite{Deffayet:2010qz} and describes  three different cosmologies, since zeros of 
  $j(\chi)\sim \sqrt{x}y\sim {a}^{-1/3}$ 
  correspond to the infinite universe size. The rightmost part of the diagram where $j(\chi)\geq 0$ describes the 
  stable branch. }
  
  \bbl{
The main accent Ref.\cite{Deffayet:2010qz} was made on the analysis of the currentless solutions (called fantom attractors) 
in the presence of a generic external matter. Although this goes beyond the scope of our program, 
we can recover 
  the same results.  Replacing in \eqref{Fr}  $\Lambda\to \rho_{ext}$ and in \eqref{FF} 
  $\Lambda\to -p_{ext}$ with $p_{ext}=w_{ext}\,\rho_{ext}$, yields 
\be      \label{Fr1}
3H^2&=&\frac32\,\alpha\PPsi^3 H-\frac{1}{4}\,\beta\,\PPsi^2
+\frac{5\eta }{24}\alpha^2 \PPsi^6+\rho_{ext}\equiv \rho_X+\rho_{ext},  \\
2\dot{H}+3H^2&=&\frac12\,\alpha\,\PPsi^2\dot{\PPsi}
+\frac14\,\beta\,\PPsi^2-\frac{\eta }{24}\,\alpha^2\, \PPsi^6-p_{ext}\equiv -p_X-p_{ext}\,,
\label{FF1}
\ee
where  $\eta$ distinguishes between theories \eqref{kgb1} and \eqref{kgb2} 
by taking values 
$0$ and $1$, respectively. 
This defines 
$\Omega_X={\rho_X}/{3H^2}$, $\Omega_e={\rho}/{3H^2}$ and $w_X={p_X}/{\rho_X}$, 
where  $H$ is obtained by setting the current  to zero (as in \eqref{HH}): 
\be            \label{HH1}
H=-\frac{\eta}{6}\,\alpha\PPsi^3 +\frac{\beta}{3\alpha\PPsi}. 
\ee
Injecting this to \eqref{Fr1} and \eqref{FF1} yields algebraic relations which determine $\psi$ 
and $\dot{\psi}$ in terms of $\rho_{ext}$ and $p_{ext}$. Computing then $\Omega_X$ and $w_X$ we recover 
the results of Ref.\cite{Deffayet:2010qz}. For example, we see  that at early times the external matter 
dominates and $\Omega_X\ll 1$ while the effective equation of state of the scalar is then 
determined by $1+w_X=-(1+w_{ext})$, whereas  at late times $\Omega_X\to 1$ and $w_X\to -1$.
}

\bbl{
Bounces in the presence of a ``hot matter" were studied in Ref.\cite{Easson:2011zy}.
Interestingly, solutions whose evolution at the turnaround point is stable were detected, 
although they still show instability somewhere \cite{Easson:2011zy}. Our bounces are 
also unstable,  in addition we do not find solutions whose evolution at the turnaround point would be stable. 
The difference must be due to the fact that   our bounces are ``cold" and not ``hot",
and also because the model considered in  \cite{Easson:2011zy} corresponds to\footnote{\bbl{After the replacement 
$X\to -X$ due to the signature change.}}  
$K=\pm X+X^3$, which 
is different from our model \eqref{kgb2}
with the negative coefficient in front of $X^3$. 
}

\section{More general Horndeski models \label{SMore}}

We have studied up to now the Palatini  version of the Horndeski models respecting  the  condition \eqref{KGB}. 
These theories are described by second order equations and are therefore free of the Ostrogradsky ghost. 
It turns out that relaxing the  condition \eqref{KGB} 
invariably produces higher derivatives within the Palatini approach. However, the 
ghost does not always arise. 
To illustrate this, let us consider a simple example obtained  by setting 
 in \eqref{horn} 
\be
G_2=G_3=0,~~~~G_4=\sigma,~~~~G_5=-\xi\, \phi, 
\ee
with constant $\sigma,\xi$. The Horndeski Lagrangian reduces to 
\be                  \label{action1}
L_{\rm H}&=&\left(\sigma R- \xi\,\phi \,G_{\mu\nu}\nabla^\mu\nabla^\nu\phi\right)\sqrt{-g}\, \nn \\
&=&\left(\sigma R+ \xi\,G_{\mu\nu}\nabla^\mu\phi\nabla^\nu\phi\right)\sqrt{-g}
+\xi\phi\nabla^\mu(G_{\mu\nu}\sqrt{-g})\nabla^\nu\phi +\ldots  \nn \\
&=&\left(\sigma R+ \xi\,G_{\mu\nu}\nabla^\mu\phi\nabla^\nu\phi\right)\sqrt{-g}+\ldots
\ee
where the dots denote total derivatives. The term $\nabla^\mu(G_{\mu\nu}\sqrt{-g})$ 
in the second line vanishes,  but it would be proportional to the non-metricity within the Palatini 
approach, hence  dropping this term is equivalent to 
 choosing a non-zero $\Delta L_{\rm P}$ in \eqref{LLLP}\footnote{\bbl{Keeping the $\nabla^\mu(G_{\mu\nu}\sqrt{-g})$ 
 term in the metric-affine approach would render the torsion dynamical, as explained  after Eq.\eqref{KGB2}.}}. 
Consider the metric-affine version of the third line in \eqref{action1}, 
\be             \label{LLL}
L_{\rm P}=\left(\sigma \oR+ \xi\,G_{\mu\nu}\,\partial^\mu\phi\partial^\nu\phi\right)\sqrt{-g}\,,
\ee
where  $\oR=g^{\mu\nu}\oR_{\mu\nu}$ and 
 $G_{\mu\nu}=\oR_{\mu\nu}-\frac12 \oR g_{\mu\nu}$.  
 Varying this with respect to $\phi$ and using \eqref{DIV} yields 
\be                     \label{eqH}
\nabla^\mu\left ( G_{(\mu\nu)}\,\partial^\nu\phi\right)
=0.
\ee
In the metric case one has 
$\onabla_\sigma g_{\mu\nu}=0$  
and $\nabla^\mu G_{\mu\nu}=0$ hence  the equation reduces  to  
$
G_{\mu\nu}\nabla^\mu\nabla^\nu\phi=0
$
which contains  only second derivatives. However,  if 
$\onabla_\sigma g_{\mu\nu}\neq 0$  
then  $\nabla^\mu G_{\mu\nu}\neq 0$ and 
the equation contains higher derivatives, which can be seen as follows.
 The Lagrangian can be represented as 
\be
L_{\rm P}=\oR_{\mu\nu}\, H^{\mu\nu}\sqrt{-g}
\ee
with 
\be
H^{\mu\nu}=\left(\sigma-\xi X\right)g^{\mu\nu}+\xi\, \partial^\mu\phi\partial^\nu\phi\,,
\ee
where as usual $X=\frac12 (\partial\phi)^2$. 
Introducing $h_{\mu\nu}$ defined by the relation 
\be
H^{\mu\nu}\sqrt{-g}=h^{\mu\nu}\sqrt{-h}\,,
\ee
hence 
\be                    \label{hhh}
h_{\mu\nu}=\sqrt{\sigma^2-\xi^2 X^2}\left(g_{\mu\nu}-\frac{\xi}{\sigma +\xi X}\,\partial_\mu\phi\partial_\nu\phi\right),
\ee
the Lagrangian becomes 
\be                \label{LLL1}
L_{\rm P}=\oR_{\mu\nu}\, h^{\mu\nu}\sqrt{-h}.
\ee
It is well-known that varying this Lagrangian with respect to the connection  yields 
\be
\Gamma^\mu_{\alpha\beta}=\frac12 \,h^{\mu\nu}\left( \partial_\alpha h_{\nu\beta}+\partial_\beta h_{\nu\alpha} 
-\partial_\nu h_{\alpha\beta}
 \right),
\ee
hence  $\Gamma^\mu_{\alpha\beta}$ is the Levi-Civita  connection for the effective metric $h_{\mu\nu}$.
Since the latter contains derivatives in \eqref{hhh}, 
it follows that $\Gamma^\mu_{\alpha\beta}$ contains second derivatives hence both $G_{\mu\nu}$ 
and the equation contains third derivatives of $\phi$.

At the same time, the relation \eqref{hhh} between 
$g_{\mu\nu}$ and $h_{\mu\nu}$ is an invertible disformal  transformation, hence one can 
consider $h_{\mu\nu}$, $\phi$ as independent variables instead of $g_{\mu\nu}$, $\phi$. 
Varying the Lagrangian with respect to $h_{\mu\nu}$ yields 
\be
R_{\mu\nu}=0, 
\ee 
which are   the vacuum Einstein equation for 
the Ricci tensor constructed from the metric $h_{\mu\nu}$ in the standard way. They imply that 
$G_{\mu\nu}=0$, hence  the scalar field equation \eqref{eqH} is fulfilled as well. Therefore, the 
theory \eqref{LLL} is simply the vacuum General Relativity for the effective metric $h_{\mu\nu}$
so that  the ghost is absent. 

The original metric $g_{\mu\nu}$ is obtained from $h_{\mu\nu}$ by inverting the relation \eqref{hhh},
and since the latter contains the scalar field $\phi$ remaining  undefined, there are infinitely many 
metrics $g_{\mu\nu}$ for a given Ricci-flat $h_{\mu\nu}$. 
This ambiguity can be  removed   by adding $
K(X,\phi)\sqrt{-g}
$
to the Lagrangian  
to produce  a non-trivial condition for $\phi$. The equations 
will still contain higher derivatives when expressed in terms of $g_{\mu\nu},\phi$, but they become second order 
equations when expressed in $h_{\mu\nu},\phi$ variables.

Summarizing, the theory \eqref{LLL} contains higher derivatives 
when parameterized in terms of $g_{\mu\nu}$ and $\phi$ hence it is outside the Horndeski family. 
At the same time, it is ghost-free   
since  the disformal transformation \eqref{hhh} removes 
the higher  derivatives, hence  it must belong to the DHOST family (similar examples were 
considered  in  \cite{Galtsov:2018xuc}).

However, in the generic case the theory turns out to be outside the DHOST family and shows ghost. 
Consider, for example, the Palatini version of the entire piece
of the Horndeski Lagrangian \eqref{horn} generated by $G_4(X,\phi)$,
\be                         \label{horn1}
L_{\rm P}&=&\left(G_4({X},\phi)\,\oR-\partial_X G_4(X,\phi)\,
\left( [\hat{\Phi}]^2- [\hat{\Phi}^2]  \right) \right)\sqrt{-g}\,. 
\ee
Solving the equation for the connection gives
\be             \label{conn}
\Gamma^\alpha_{\mu\nu}=\left\{^\alpha_{\mu\nu}\right\}
+D^\alpha_{\mu\nu}
\ee
where $D^\alpha_{\mu\nu}$ is displayed  in the Appendix. 
Injecting this back to $L_{\rm P}$ yields for a generic $G_4(X,\phi)$  a metric Lagrangian that  belongs 
neither to the Horndeski nor to DHOST family. 
Therefore the theory contains ghost. For the particular choice $G_4(X,\phi)=f(\phi)X$
the theory can be shown to be of the DHOST type, but it is unclear if the ghost 
can be removed in other cases, for example by adding to the Lagrangian 
a non-trivial $\Delta L_{\rm P}$ as in \eqref{LLLP}. 
 
 \bbl{The Palatini versions of the parts 
of the  Lagrangian \eqref{horn}  containing  $G_5(X,\phi)$ remains totally unexplored 
since the torsion is then dynamical. }

\section{Concluding remarks \label{SFin}}

Summarizing the above discussion, we have studied what happens if the Horndeski theory is treated within the Palatini approach. 
It turns out that there are infinitely many metric-affine versions $L_{\rm P}$ of the original Horndeski Lagrangian which differ from 
each other by terms proportional to the non-metricity tensor, as  expressed by \eqref{LLLP}. 
Each $L_{\rm P}$ defines a theory which is equivalent to a certain metric theory with the Lagrangian 
obtained by injecting the algebraic solution for the connection back to $L_{\rm P}$. Therefore, the metric-affine generalisations 
of the  Horndeski theory reduce  again to metric theories for a gravity-coupled scalar field. 

Every such a metric theory can either belong to the original Horndeski family,
or it can be of a more general DHOST type, or it can be something else, in which case it has the Ostrogradsky ghost. 
Therefore, the metric-affine generalisations of the Horndeski theory can be ghost-free but not all of them are  ghost-free.

It is interesting to know when these theories are ghost-free.  We were able to give the answer for the KGB 
subset of the Horndeski theory defined by the condition \eqref{KGB}: it turns out that its metric-affine version defined by 
\eqref{RG}--\eqref{SP} is ghost-free because it yields  a theory which is again in the metric KGB class. 
We have also checked that its generalisation defined by \eqref{LLLP}, where $\Delta L_{\rm P}$ contains only 
the linear in the non-metricity terms  shown in  \eqref{LLP}  remains ghost-free   \cite{PREP1}. 
\bbl{We classified all homogeneous and isotropic  cosmologies in  these theories. } 

The situation with more general Horndeski models is more complicated and will be reported separately \cite{PREP}. 
It is possible  that the metric-affine versions of the parts 
of the Horndeski Lagrangian containing $G_4(X,\phi)$ could  be made ghost-free by carefully adjusting  $\Delta L_{\rm P}$
but is unclear  if this procedure works for generic $G_4(X,\phi)$. 
The situation is totally unexplored if  the Lagrangian contains  $G_5(X,\phi)$.

One should also say that the Horndeski theory is not the only one whose  
metric-affine versions can be ghost-free. For example, 
the theory described 
\be
S_{\rm P}[\Gamma^\sigma_{\alpha\beta},g_{\mu\nu},\phi]=\int\left(\oR_{\mu\nu}\left[ G_4(X,\phi)g^{\mu\nu}+ 
G_5(X,\phi)\,\partial^\mu \phi \partial^\nu\phi \right]+K(X,\phi)\right)\sqrt{-g}\, d^4 x~~~~~
\ee
has second order equations but does not reduce to Horndeski theory when the non-metricity vanishes. 

Another example is provided by the Lagrangian \cite{Aoki:2018lwx}
\be                         \label{horn2}
L_{\rm P}&=& \left\{ K(X,\phi)+G_3(X,\phi)[\hat{\Phi}] 
+G_4({X},\phi)\,\oR
-\partial_X G_4(X,\phi)\,
\left( [\hat{\Phi}]^2- [\hat{\Phi}^2]  \right) \right.  \nn \\
&&- \left.\frac{ \partial_X G_4(X,\phi) }{X}\,(\nabla_\mu  X-[\hat{\Phi}]\nabla_\mu\phi )\nabla^\mu X \right\}\sqrt{-g}\,, 
\ee
where   the terms in the first line are the same as in the Horndeski theory, whereas  those 
in the second line do not have the Horndeski structure. Adding to this 
 a 
suitably chosen $\Delta L_{\rm P}$ made of the non-metricity  and varying yields a particular member of the DHOST family \cite{Aoki:2018lwx}
(we were able to confirm this \cite{PREP}),
hence the theory is ghost-free.

An example of a completely different  type  is provided by the Born-Infeld theory,
 \be
S_{\rm P}[\Gamma^\sigma_{\alpha\beta},g_{\mu\nu},\phi]=\int\left(\sqrt{-\det\left(g_{\mu\nu}+\sigma \oR_{(\mu\nu)} \right) }+K(X,\phi)\sqrt{-g}\right) d^4 x,~~~~
\ee
which has   second  order equation
\cite{BeltranJimenez:2017doy}.  It follows that  the Horndeski Lagrangian is not the most general one that 
leads to second order field equations within the Palatini approach. An interesting problem would be to 
 find the most general ghost-free metric-affine theory.

{\bf Acknowledgments.--}  It is a pleasure to acknowledge discussions with Evgeny Babichev, Thibault Damour, Alexander Vikman, 
 and especially with Dr. Katsuki Aoki who explained to us his work and helped to clarify a number of important 
issues.  M.S.V. thanks for hospitality the YITP in Kyoto, where a part of this work was completed. 
His work was also partly supported by the  PRC CNRS/RFBR project grant, as well as 
by the Russian Government Program of Competitive Growth 
of the Kazan Federal University. 

\section*{Appendix}

\setcounter{section}{0}
\setcounter{equation}{0}
\setcounter{subsection}{0}

\renewcommand{\theequation}{A.\arabic{equation}} 

\renewcommand{\theequation}{A.\arabic{equation}}

Here is the explicit form of the non-metric part of the connection  in \eqref{conn}:
\begin{align*}
{D}^\alpha_{\mu\nu}&=-2A\nabla^\alpha\nabla_{(\mu}\phi\nabla_{\nu)}\phi+A\nabla^{\alpha}\phi\nabla_\mu\nabla_\nu\phi+B\delta^{\alpha}_{(\mu}\nabla_{\nu)}\nabla_\beta\phi\nabla^\beta\phi\\
&+B g_{\mu\nu}\nabla^\alpha\nabla^\beta\phi\nabla_\beta\phi\left(\dfrac{3}{2}-\dfrac{2XG_{4X}}{G_4}\right)-AB\nabla_\mu\phi\nabla_\nu\phi\nabla^\alpha\nabla^\beta\phi\nabla_\beta\phi\left(5-\dfrac{6XG_{4X}}{G_4}\right)\\
&+g_{\mu\nu}\nabla^\alpha\phi\Bigg(\dfrac{AC}{6}\dfrac{G_4}{G_{4X}}\left(14XG_{4X}-3G_4\right)\Box\phi  \\
&+\dfrac{ABC}{3G_4}X\left(7G_4^2-42G_4XG_{4X}+48X^2G_{4X}^2\right)Y\\
&-\dfrac{G_{4\phi}C}{2G_{4X}}\left(G_4-2XG_{4X}\right)\Bigg)+\nabla^\alpha\phi\nabla_\mu\phi\nabla_\nu\phi\Bigg(-\dfrac{AC}{3}\left(G_4+12XG_{4X}\right)\Box\phi\\
&-\dfrac{ABC}{3G_4}\left(7G_4^2-42G_4XG_{4X}+48X^2G_{4X}^2\right)Y+\dfrac{G_{4\phi}}{G_4}C\left(G_4-4XG_{4X}\right)\Bigg)\\
&+2\Bigg(\dfrac{AC}{6G_{4X}}\left(G_4+12XG_{4X}\right)\left(2XG_{4X}-G_4\right)\Box\phi\\
&+\dfrac{ABC}{3G_4G_{4X}}\left(4G_4^3-11G_4^2XG_{4X}+30G_4X^2G_{4X}^2-24X^3G_{4X}^3\right)Y\\
&+\dfrac{G_{4\phi}}{2G_4}\dfrac{C}{G_{4X}}\left(G_4-2XG_{4X}\right)\left(G_4-4XG_{4X}\right)\Bigg)\delta^{\alpha}_{(\mu}\nabla_{\nu)}\phi, \label{ConnnectionSol}
\end{align*}
with $G_{4X}= \partial_X G_4(X,\phi)$ and the functions $Y,A,B,C$ defined as 
\begin{align}
&Y= \nabla_\alpha\phi\nabla_\beta\phi\nabla^\alpha\nabla^\beta\phi,~~~~~
A= \dfrac{G_{4X}}{G_4+2XG_{4X}},\nn \\
&B= \dfrac{G_{4X}}{3G_4-2XG_{4X}},~~~~~~
C=\dfrac{G_{4X}}{G_4^2-4G_{4}XG_{4X}+8X^2G_{4X}^2}. \nn
\end{align}
The function $A,B$ here should not be confused with those used in the main text.



\providecommand{\href}[2]{#2}\begingroup\raggedright\endgroup

\end{document}